\documentclass[%
 aip,
 amsmath,amssymb,
 reprint,%
]{revtex4-1}

\usepackage{graphicx}% Include figure files
\usepackage{dcolumn}% Align table columns on decimal point
\usepackage{bm}% bold math
\usepackage[utf8]{inputenc}
\usepackage[T1]{fontenc}
\usepackage{mathptmx}
\usepackage{etoolbox}
\usepackage{amsmath,amssymb,amsthm,color,dsfont,graphicx,mathtools,soul,xcolor,fullpage}

\definecolor{mygray}{gray}{0.3}
\newcommand{\edits}[1]{{\color{black}{#1}}}

\makeatletter
\def\@email#1#2{%
 \endgroup
 \patchcmd{\titleblock@produce}
  {\frontmatter@RRAPformat}
  {\frontmatter@RRAPformat{\produce@RRAP{*#1\href{mailto:#2}{#2}}}\frontmatter@RRAPformat}
  {}{}
}%
\makeatother

\begin{document}

\preprint{AIP/123-QED}

\title[]{Computing transition path theory quantities with trajectory stratification}
% Force line breaks with \\
\author{Bodhi P. Vani}
\affiliation{ 
Department of Chemistry and James Franck Institute, University of Chicago, Chicago IL 60637}
\author{Jonathan Weare}%
\affiliation{ 
Courant Institute of Mathematical Sciences, New York University, New York 10012
}%

\author{Aaron R. Dinner$^*$}
\affiliation{ 
Department of Chemistry and James Franck Institute, University of Chicago, Chicago IL 60637}
\email{dinner@uchicago.edu}

\date{\today}% It is always \today, today,

\begin{abstract}
%Transition path theory enables computation of statistics that characterize reaction probabilities and pathways.
Transition path theory computes statistics from ensembles of reactive trajectories.  
A common strategy for sampling reactive trajectories is to control the branching and pruning of trajectories so as to enhance the sampling of low probability segments.
However, it can be challenging to apply transition path theory to data from such methods because determining whether configurations and trajectory segments are part of reactive trajectories requires looking backward and forward in time.
Here, we show how this issue can be overcome efficiently by introducing simple data structures.  We illustrate the approach in the context of nonequilibrium umbrella sampling (NEUS), but the strategy is general and can be used to obtain transition path theory statistics from other methods that sample segments of unbiased trajectories.
\end{abstract}

\maketitle

\section{Introduction}

%Molecular dynamics is good, important. Why we want to compute rates.

%We want to characterize mechanisms and we want to understand statistical significance of mechanisms in multi-pathway reactions.

One of the main purposes of simulations is to learn how processes occur. 
However, most processes of interest occur on timescales that are orders of magnitude longer than the numerical integration time step.  In the case of molecular systems, on which we focus here, the time step for all-atom models is in the femtosecond range, while conformational transitions of functional significance typically take place in the microsecond to seconds range.  As a result, at best, only a small number of events can be observed by direct simulation. Moreover, even if one obtains examples of events of interest, the number of dynamical variables is generally sufficiently large that it is not obvious which are the key steps \cite{ma_automatic_2005}, leave alone their likelihoods and the distributions of times that they occur.  
 
Transition Path Theory (TPT) provides a means of computing these statistics for transitions between two metastable states from the ensemble of reactive trajectories (i.e., those that contain an event)  \cite{Vanden-Eijnden2006,e_towards_2006,metzner_illustration_2006}.  A key idea in TPT is that these statistics can be expressed as products over the steady-state distribution and the probabilities of coming from or going to the metastable states (known as commitment probabilities or committors).  This factorization allows estimating these probabilities separately, which opens the door to computing statistics of reactive trajectories from a broader range of data.

Markov State Models (MSMs) are a popular means of computing TPT statistics \cite{noe2009constructing,bowman2013introduction}.  
In MSMs, dynamics are coarse-grained to transitions between a set of discrete states, the probabilities of which are assumed to be independent of the sequence of states visited.  The advantage relative to direct simulation comes from the fact that there is considerable freedom in how one estimates the state-to-state transition probabilities \cite{wu_variational_2017,buchete2008coarse}.  Dynamical Galerkin Approximation (DGA) \cite{thiede_galerkin_2019,strahan2021long} is a generalization of MSMs for TPT statistics that accounts for the boundary conditions of the statistics \cite{russo2021unbiased} and represents them through expansions over sets of basis functions (which can take a form beyond indicator functions on discrete states).  
Recently, TPT statistics have also been computed with milestoning \cite{elber2017calculating}, which instead assumes that the system reaches an equilibrium within selected parts of the configuration space (the milestones) \cite{faradjian2004computing,elber2020milestoning}.

Although often reasonable, the  above approximation schemes can break down, and, even if they do not, they ultimately limit the resolution with which statistics can be computed.  Methods that sample the ensemble of trajectories that connect the metastable states without such assumptions can in principle provide results with arbitrary resolution.  One such method is Transition Path Sampling (TPS),  a Monte Carlo procedure that accepts and rejects whole trajectories \cite{dellago1998transition,dellago1998efficient,dellago2002transition}.  However, generally it is more efficient to divide the configuration space and sample segments of trajectories that transition between regions.  This idea is at the foundation of the Weighted Ensemble (WE) \cite{huber_weighted-ensemble_1996,zhang2007efficient,zuckerman2017weighted}, Transition Interface Sampling (TIS) \cite{bolhuis2003transition,van2003novel}, Forward Flux Sampling (FFS) \cite{allen2005sampling,allen2009forward}, Nonequilibrium Umbrella Sampling (NEUS) \cite{warmflash2007umbrella,dickson2009separating,vanden2009exact,dickson2010enhanced,dinner2020stratification}, and Exact Milestoning (EM) \cite{bello2015exact} methods.  With the exception of TIS, these methods allow treatment of microscopically irreversible dynamics, which is also not straightforward in TPS.

The differences between these methods are in the details of the trajectory segments they select and how they track their probabilities to enable reconstruction of the overall statistics.  Generally the interfaces are specified through collective variables that combine information from multiple coordinates used for the underlying numerical integration.  TIS and FFS require non-intersecting interfaces because they are formulated in terms of the probabilities of going from one to another in order.  WE, NEUS, and EM do not have this restriction and thus readily allow control of sampling in more than one collective variable.  In its traditional form, WE tracked the weight of each trajectory segment (equivalent to a member of the ensemble) individually.  NEUS instead tracked the weight of a subpopulation within a region (termed a stratum in statistics) and redistributed the weight to satisfy a global flux balance condition.  Later implementations of WE that incorporate this idea \cite{bhatt2010steady} are very similar to NEUS, as is EM, which relaxes the distributions on the milestones.  Several of these algorithms can thus be described by a common framework, trajectory stratification \cite{dinner2018trajectory}.  

Because these methods sample trajectory segments rather than whole trajectories, care must be used to weight trajectory segments appropriately when computing statistics.  Vanden-Eijnden and Venturoli  \cite{vanden2009exact} showed that the rate can be obtained exactly from NEUS by augmenting the dynamics with information about the last metastable state visited, and this scheme was further refined and applied in ref.\ \onlinecite{dickson2009separating}.  However, the rate requires only tracking transitions into the metastable states, making it easier to compute than statistics that require information about intervening states.  Vanden-Eijnden and Venturoli \cite{vanden2009exact} provided a formula for computing the (backward) committor from the region weights, but, to the best of our knowledge, statistics such as committors and probability currents of reactive trajectories (reactive currents for short), which reveal microscopic likelihoods of reaction and reaction pathways, respectively, have not been computed from such methods.   The challenge presented by these quantities is that each trajectory segment can contribute to an infinite number of complete trajectories, and determining the ones that contribute to a given statistic requires looking both forward and backward in time.

Here, we introduce a simple bookkeeping procedure and practical formulas for computing committors and reactive currents.  While we present our methods in terms of the NEUS algorithm, they can be adapted to the other algorithms for sampling trajectory segments described above.  The paper is organized as follows.  In Section \ref{kineticstatistics}, we define these quantities in the TPT framework, and in an associated appendix we provide an explicit justification for an estimator of the current.  In Section \ref{NEUS} we provide a succint description of the NEUS algorithm and show how it must be modified.  We demonstrate the algorithm for isomerization of a peptide in Section \ref{numerical}.

\section{Transition path theory}
\label{kineticstatistics}
We are interested in learning the statistics of reactive trajectories.  In TPT, these are defined to be trajectories that start in a state $A$ and end in a non-overlapping state $B$ without returning to $A$.  To delimit their ends, we define
\begin{align}
t_+(t)&= \min( t' > t,\   X_{t'} \in A \cup B) \\
t_-(t)&= \max( t' < t,\  X_{t'} \in A \cup B),
\end{align}
where $X_{t'}$ is the configuration at time $t'$, and $X_t\in (A\cup B)^c$.
The forward committor $q_+(x)$ is the probability that the system goes to $B$ before $A$ starting from state $x$:
\begin{equation}
    q_+(x) = {\bf P}[X(t_+(0))\in B|X(0)=x].
\end{equation}
Similarly, the backward committor $q_-(x)$ is the probability that a system at $x$ came from $A$ after $B$:
\begin{equation}
    q_-(x) = {\bf P}[X(t_-(0))\in A|X(0)=x].
\end{equation}
In the present study, for visualization, we project the committors onto the space defined by a vector of collective variables, $\theta(x)$, weighted by $\pi(x)$:
\begin{equation}
    q_\pm^\theta(\Theta) =\int q_\pm(x)\pi(x)\delta(\theta(x)-\Theta)dx,
\end{equation}

A key element of TPT is that the probability that a trajectory is reactive can be expressed in terms of quantities that are local in space.  Specifically, the probability that a trajectory that passes through $x$ is reactive is proportional to $\pi(x)q_-(x)q_+(x)$, where $\pi(x)$ is the steady-state distribution.
The fact that TPT is based on local statistics aids in interpretation, but it also can make obtaining certain statistics challenging.  We discuss how TPT of an augmented process can be used to obtain additional statistics that require knowledge of sequences of states visited in ref.\  \onlinecite{lorpaiboonATPT}. Of course, the analysis we propose here can be complemented by direct analysis of the reactive trajectories as well. 

\subsection{Reactive current}

The reactive current is a central quantity in TPT \cite{Vanden-Eijnden2006,e_towards_2006,metzner_transition_2009}.  It can be used to define reactive pathways, and averages over reactive trajectories---in particular the rate---can be computed from it.  The reactive current is defined by considering a surface $S$ that divides the space into two regions:  $C$ containing $A$ and its complement $C^c$ containing $B$.  Then, the reactive current, $\edits{I}_{AB}(x)$, is the vector field whose integral is the reactive flux across $S$:
\begin{multline}
\label{defJ}
\int_{S}\edits{I}_{AB} \cdot \hat{n}_S d\sigma_S =
\lim_{\tau \to 0}\lim_{T\to \infty}\frac{1}{2T\tau}\int_{-T}^T[\mathbf{1}_{C}(X(t))\mathbf{1}_{C^c}(X(t+\tau))\\
- \mathbf{1}_{C^c}(X(t))\mathbf{1}_{C}(X(t+\tau))]
\mathbf{1}_A(X(t_-(t)))\mathbf{1}_B(X(t_+(t+\tau)))dt
\end{multline}
where $\hat{n}_S$ is a unit vector normal to $S$ (pointing from $C$ to $C^c$), $d\sigma_S$ is a surface element, $\tau$ is a lag time, and 
\begin{equation}
\mathbf{1}_{D}(x) =
    \begin{cases}
    1 & \text{if } x\in D \\
    0 & \text{otherwise}.
    \end{cases} 
\end{equation}

In general, the full state space of the system has many dimensions.   
$\edits{I}_{AB}(x)$ is a vector field in the full space of dynamical variables, of which there are many in general.  However, most of these are irrelevant for determining whether the reaction proceeds, and we seek a small number of key dimensions.  Because these are often combinations of the original dynamical variables, we term them collective variables.  We show in the Supplemental Information that 
\begin{multline}\label{JABtheta}
\edits{I}_{AB}\cdot\nabla \theta(x) =\pi(x) q_-(x)q_+(x) \\
\times\lim_{\tau\to 0}\frac{1}{2\tau} 
{\bf E}[(\theta(X(\tau))-\theta(X(-\tau))|\\
X(t_-(0))\in A , X(t_+(0))\in B, X(0)=x].
\end{multline}
We use $\bf E$ to denote expectations over the steady-state trajectory ensemble. This formula is simply the product of the  probability of being on a reactive trajectory and the average increment per unit time of $\theta$ along a reactive trajectory.
We project this current onto the space of collective variables:
\begin{align}\label{JABprojection}
\edits{I}_{AB}^\theta(\Theta) &= \int \edits{I}_{AB}(x)\cdot \nabla \theta(x) \delta(\theta(x)-\Theta)dx.
\\&= 
\lim_{\lvert d\Theta\rvert \rightarrow 0} \frac{1}{\lvert d\Theta\rvert}\int_{\{\theta(x)\in d\Theta\}}  \edits{I}_{AB}(x)\cdot \nabla \theta(x)dx.
\end{align}
We recently showed that this projected reactive current can also be used to compute the reactive flux through an integral similar in form to \eqref{defJ} but in the $\theta$ space; the projected current $\edits{I}_{AB}^\theta(\Theta)$ can in turn be used to compute the rate \cite{strahan2021long}.
The expressions in \eqref{JABtheta} and \eqref{JABprojection} allow us to obtain the reactive current in the space of collective variables from reactive trajectories. 

In the next section, we discuss how reactive trajectories can be sampled efficiently.
Although we treat time as continuous above, with a view toward developing a practical algorithm, we henceforth assume a dynamics with discrete time steps; the lag time $\tau$ is thus an integer multiple of the time step in practice.

\section{Trajectory stratification}
\label{NEUS}

% Aaron to integrate
The idea of stratification in general is to obtain better overall statistics by controlling the sampling of subpopulations (strata).
The most familiar form of stratification in molecular simulations is umbrella sampling \cite{torrie1977nonphysical,pangali1979monte,frenkel2001understanding,thiede2016eigenvector,dinner2020stratification}, which is frequently used to compute free energies. 
In this case, the strata are defined by regions in a space of collective variables.  Each of the strata is sampled independently by a copy of the system, and the information is then combined to obtain unbiased averages. 
Trajectory stratification extends this strategy from states to trajectories and thus enables treating microscopically irreversible systems and estimating dynamical statistics (for microscopically reversible or irreversible systems) \cite{warmflash2007umbrella,dickson2009separating,vanden2009exact,dickson2010enhanced,dinner2020stratification}.

Here, we focus on the case of steady-state, time-independent quantities.   
%The essential idea of trajectory stratification \cite{warmflash2007umbrella,dickson2010enhanced,dinner2018trajectory} is that expectations are sums over strata.
%We sample the strata in parallel by running unbiased dynamics between their boundaries and detect transitions between strata by associating with $X(t)$ a variable $J(t)$ that reports the index of the stratum at time $t$.  
To formulate the algorithm mathematically, we  associate with $X(t)$ a variable $J(t)$ that reports the index of the stratum at time $t$. 
%(readers should not confuse the index process $J(t)$ with the currents $\edits{I}_{AB}(x)$ and $\edits{I}_{AB}^\theta(\Theta)$). 
For example, a trajectory that spent one time step in stratum 1, two time steps in stratum 2, went back to stratum 1 for one time step, and then spent three time steps in stratum 5 would have $J(t=1) = 1$, $J(t=2)=2$, $J(t=3)=2$, $J(t=4)=1$, $J(t=5)=5$, $J(t=6)=5$, and $J(t=7)=5$.
Then, denoting the time that a trajectory first exits a stratum as $s=\min\{t:J(t)\neq J(0)\}$, we can write expectations as
\begin{equation}
    {\bf E}[f] = \frac{1}{\Omega}\sum_iz_if_i,
\end{equation}
where  $\Omega$ is a normalization factor and
\begin{equation}\label{fieqn}
    f_i=
    \int_x \sum_{t} f(X(t)) {\bf P}_{x,i} [X(t), t<s]  \pi_i(dx).
\end{equation}
${\bf P}_{x,i}$ indicates that we are conditioning the probability on starting at configuration $x$ in stratum $i$;
$\pi_i(dx)\equiv\pi_i(x)dx$ is the probability of being in the differential volume element $dx$ conditioned on entering stratum $i$ at that state, and $z_i$ is the normalization factor for $\pi_i(dx)$.
Operationally, \eqref{fieqn} corresponds to initiating trajectories from steady-state entry points to stratum $i$, terminating them when they leave the stratum, and averaging over the sampled points.
We show in ref.\ \onlinecite{dinner2018trajectory} that, for ergodic averages, the vector $z$ can be obtained by solving an eigenequation:
\begin{equation}\label{affineeig3}
{z}^\text{\tiny T} G =  {z}^\text{\tiny T}.
\end{equation}
The matrix $G$ tracks the number of transitions between strata; we define it precisely below in \eqref{eq:Gdefn}.
%, which we detect by associating with $X(t)$ a variable $J(t)$ that reports the index of the stratum at time $t$.  
Physically, 
%$z$ can be viewed as the probability flux into a stratum, and 
\eqref{affineeig3} can be viewed as a statement of global flux balance.

The key idea in the present work is that we define the index process so as to track the last metastable state visited, in the spirit of refs.\ \onlinecite{dickson2009separating} and \onlinecite{vanden2009exact}.  In other words, if there are $K$ strata with domains $D_j$ defined in terms of a set of order parameters, then $J(t)$ runs from 1 to $2K$:
\begin{equation}\label{jprocess}
    J(t)=K \mathbf{1}[X(t_-(t))\in B] + \sum_{j=1}^{K}j\mathbf{1}[X(t)\in D_j].
\end{equation}
Thus strata with indices from $1$ to $K$ capture data for trajectories that last came from state $A$, and strata with indices $K+1$ to $2K$ represent trajectories that last came from state $B$.

\subsection{A self-consistent iteration}

Given the index process, we can now define $z_i$, $\pi_i$, and $G_{ij}$ more precisely (we note that these quantities correspond to the steady-state averages of $\bar{z}_i$, $\bar{\pi}_i$, and $\bar{G}_{ij}$ defined in (24), (31), and (34), respectively in ref.\ \onlinecite{dinner2018trajectory}; we suppress the overbars here for simplicity of notation):
\begin{equation}\label{eqn:pi_i2}
    \pi_i(dx) =\lim_{T\to\infty}\frac{1}{T} \frac{\sum_{\edits{t=0}}^{T}{\bf P}[J(t) = i, J(t-1)\neq i, X(t)\in dx]}{z_i}
\end{equation}
and
\begin{equation}
    z_i = \lim_{T\to\infty}\frac{1}{T}\sum_{\edits{t=0}}^T{\bf P}[J(t) = i, J(t-1) \neq i].
\end{equation}
To obtain $G_{ij}$, we draw states from $\pi_i(dx)$ and evolve them until they leave stratum $i$:
\begin{align}\label{eq:Gdefn}
    G_{ij}& = \lim_{T\to\infty}\frac{1}{T}\frac{\sum_{\edits{t=0}}^T{\bf P}[J(t)=j, J(t-1)=i]}{z_i}\\
    &=\int {\bf P}_{x,i}[J(s)=j]\pi_i(dx).
\end{align}
%\begin{equation}
%G_{ij}= \frac{  \sum_t^\infty \mathbf{P}\left[ J(t+1)=j,J(t)=i\right]}
%{\sum_t^\infty \mathbf{P}\left[J(t)=i,J(t-1)\neq i\right]},
%\end{equation}
Furthermore, we add to the distribution  of entry points to stratum $j$ that come from stratum $i$:
\begin{align}
\gamma_{ij}(dx)&=\lim_{T\to\infty}\frac{1}{T} \frac{\sum_{\edits{t=0}}^T {\bf P}[J(t)=j,J(t-1)=i,X(t)\in dx]}{G_{ij}z_i}\\
&=\int {\bf P}_{y,i}[J(s)=j,X(s)\in dx]\pi_i(dy).
\end{align}

%Operationally however, it is useful to note that we are simulating excursions after sampling from distributions of initial points at boundaries of strata. 
%\begin{equation}
%\tilde{\pi}_{j}(dx)=\frac{\sum_t^\infty P[X(t)\in dx,J(t)=j,J(t-1)\neq j]}{\tilde{z}_{Ij}}
%\end{equation}
%where the normalization factor is
%\begin{equation}
%\tilde{z}_{Ij}=\sum_t^\infty P[J(t)=j,J(t-1)\neq j]
%\end{equation}
Thus, given $\pi_i(dx)$, we can compute $G_{ij}$ and $\gamma_{ij}(dx)$, and in turn $z_i$ from \eqref{affineeig3}.  
Because the entries of $G$ depend on the dynamics within the strata, and those are initialized using $G$, the algorithm is iterative in nature.
To complete the iteration, we start from \eqref{eqn:pi_i2} and write $\pi_j(dx)$ in terms of the other quantities:
%\begin{equation}
%    \tilde{\pi}_j=\frac{1}{\tilde{z}_j}\sum_i \tilde z_i G_{ij} \gamma_{ij}
%\end{equation}
\begin{align}
        \pi_j(dx)
%        &=\lim_{t\rightarrow\infty}\frac{\mathbf{P}[J(t)=j,X(t)\in dx]}{z_j}\nonumber\\
        &=\lim_{T\rightarrow\infty}\frac{1}{T}\frac{\sum_{\edits{t=0}}^T \mathbf{P}[J(t)=j,J(t-1)\neq j,X(t)\in dx]}{z_j}\nonumber \\
        &=\lim_{T\rightarrow\infty}\frac{1}{T} \frac{\sum_{i\neq j}\sum_{\edits{t=0}}^T  \mathbf{P} \left[J(t)=j,J(t-1)=i,X(t)\in dx\right]}{z_j} \nonumber\\
        & = \frac{1}{z_j}\sum_{i\neq j} z_i G_{ij} \gamma_{ij}(dx).\label{pi_j_expr2}
\end{align}
%where in the second line we exploited time translational invariance since we are interested in ergodic averages.

%\subsection{Dynamical quantities as expectations from stratified sampling}

%From the viewpoint of an infinite trajectory, the value we are computing is
%
%\begin{align}
%    J_{AB} \cdot \nabla\theta (x) & =\lim_{T\to\infty} \frac{1}{T} \int_{-T}^T \frac{ (\theta(X(t-T)-\theta(X(t+\tau))}{2\tau} q_-(x) q_+(x) \mathbf{1}[X(t)=x] dt\\
%    &= \lim_{T\to\infty} \frac{1}{T} \int_{-T}^T \frac{ (\theta(X(t-\tau)-\theta(X(t+\tau))}{2\tau} \mathbf{1}[X(t_+_{AB}(t)\in B]\mathbf{1}[X(t_-(t)\in A]  \mathbf{1}[X(t)=x] dt
%\end{align}
%
% In terms of the NEUS stratification,
% %
% \begin{equation}
% J_{AB} \cdot \nabla\theta=\sum_i z_i \int \frac{ (\theta(X(t)-\tau)-\theta(X(t+\tau))}{2\tau} \mathbf{1}[X(t_+(x)\in B]\mathbf{1}[X(t_-(t)\in A] \mathbf{1}[J(t_+(x)-1)=i] \pi_i dt
% \end{equation}
% %

\subsection{Algorithm}
\label{secn:algorithm}
If one had a long trajectory that passed between the metastable states many times, one could compute \eqref{JABtheta} by identifying the reactive trajectories from $A$ to $B$ and accumulating the increment per unit time in $\theta$ as a function of $\theta$.  However, in the stratification scheme above, we generally do not know whether a copy of the system (random walker, or walker for short) is part of a reactive trajectory as we are evolving it.  

The key to addressing this issue is based on the fact that, 
as described previously \cite{warmflash2007umbrella,dickson2009nonequilibrium,dickson2009separating,dinner2018trajectory}, in practice, $\gamma_{ij}(dx)$ is a list of configurations that are saved when walkers enter stratum $j$ from stratum $i$.  Consequently, for each walker, we can save the $\theta$ points visited, the increments in $\theta$ at those points, and a pointer to the configuration in $\gamma_{ij}(dx)$ from which the walker originated; then, when a walker enters $A$ or $B$,
%from strata 1 to $K$ enters $B$ or a walker from strata $K+1$ to $2K$ enters $A$ (see \eqref{jprocess})
we reconstruct the full sequence of walkers connecting the metastable states and in turn the sequence of $\theta$ values visited and the associated increments, and we add these data to the appropriate sums.  
%\edits{Importantly, we do not overwrite elements of the $\gamma_{ij}(dx)$ list, as doing so may bias the statistics as discussed in ref.\ \onlinecite{dickson2009separating}.}
The algorithm is as follows.

%In essence, we want to make use of this scheme to generate ensembles of reactive trajectories. We can view this as accumulating information as the reaction progresses so that each point in the flux list contains not only information of the structure (as it must in regular NEUS) but also a pointer to its past trajectory. The true observable being recorded is an accumulated histogram of points, an accumulated estimate for $J_{AB}\cdot\nabla\theta$ for chosen $\theta$ variables. Due to our stratification scheme, each point also records which basin it originated from. The actual computation occurs only when walkers enter a basin. Instead, it is helpful to frame the entire current contribution of the reactive trajectory as an observable associated with its last point. 

%Let us assume we are computing the reactive current over a grid point ${\Theta_i}$ with indicator functions over $D_{\Theta_i}$, so that for a walker $n$ in stratum $j$ the observable is:
%
%\begin{equation}
%    \{v_{AB}(\Theta_0),v_{AB}(\Theta_1),...,v_{AB}(\Theta_N)\}_{n,j}
%\end{equation}
%
%where
%
%\begin{equation}
%    v_{AB}(\Theta_i)_{n,j}= z_j  \sum_{t=t_-(t_n)}^{t_n} \frac{ (\theta(X(t)-\tau)-\theta(X(t+\tau))}{2\tau} \mathbf{1}[X(t)\in D_{\Theta_i}]
%\end{equation}
%
%This observable is then added over walkers that enter $B$.

\begin{figure}%[h!]
%\centering
\hspace*{-1.8em}
\includegraphics[scale=0.2]{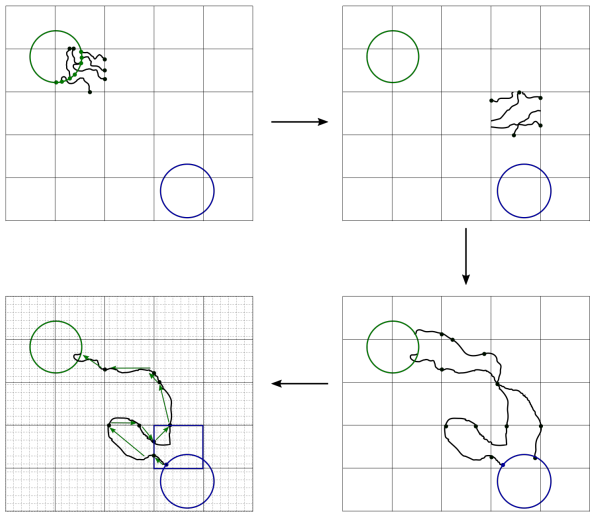}
\caption{Schematic of algorithm. 
(upper left) Define the metastable states and the strata (black grid) and initialize the data structures (step \ref{initialization}).  In the example shown, trajectories are initialized from the upper left metastable state boundary, so the last metastable state visited is known. (upper right) In each stratum, draw configurations from $\gamma_{ij}(dx)$, save pointers to the associated $\gamma_{ij}(dx)$ elements, and then run unbiased dynamics from these configurations until the trajectories exit the stratum (steps \ref{drawwalkers} and \ref{walkerexit}).  A single example stratum is shown. (lower right) As the simulation progresses, each point in $\gamma_{ij}(dx)$ may give rise to segments of reactive or unreactive trajectories.  Save the collective variable values associated with these trajectory segments (step \ref{cvsave}).
(lower left) When a reactive or unreactive trajectory is realized (i.e., a trajectory segment enters a metastable state), trace back from pointer to pointer to determine the sequences of strata and collective variable grid sites (gray grid) visited (step \ref{tptptr}).  Use these to increment unnormalized statistics with appropriate weights (steps \ref{tptsave}a-c).  Note that the collective variable grid sites and the strata need not coincide.
%(upper left) Set up a stratification scheme and initialize with configurations in $\gamma_{ij}(dx)$ (i.e., configurations that enter stratum $j$ from stratum $i$) such that we know which metastable state last visited. (upper right) In each stratum, choose a fixed number of walkers and release excursions. (lower right) As the simulation progresses, each point in $\gamma_{ij}(dx)$ may give rise to reactive or unreactive trajectories. For each flux point, save the collective variable and increment vectors of the previous trajectory, as well as a pointer to the walker from which the trajectory originated. (lower left) When a reactive or unreactive trajectory is realized (i.e., a walker enters a metastable state) trace back to connect the flux points of origin for the trajectory and add relevant values to the estimates as described in the algorithm.
\label{fig:tptschematic}
}
\end{figure}

\begin{enumerate}
    \item{Initialize simulation quantities as follows. \label{initialization}
    \begin{enumerate}
        \item Define the $2K$ strata as in \eqref{jprocess}.
        
        \item Use preliminary trajectories (obtained from either unbiased simulation or an enhanced sampling method such as STePS \cite{guttenberg2012steered}) to populate $\gamma_{ij}(dx)$. In practice, $\gamma_{ij}(dx)$ is a list of configurations used to represent the distribution of entry points to stratum $j$ from stratum $i$, and each element contains a full configuration and a pointer (set to null for the configurations from the preliminary trajectories).   In the present study, we allow the list to grow throughout the simulation (i.e., we do not overwrite configurations, as doing so may bias the statistics as discussed in ref.\ \onlinecite{dickson2009separating}).
        
        \item Estimate $G_{ij}$ as the number of transitions from stratum $i$ to stratum $j$ in the preliminary trajectories, normalized by the number of transitions out of stratum $i$.
        
        \item Initialize $z$ by solving \eqref{affineeig3}.
        
        \item Create a grid over the collective variables and denote the volume of each grid element by $\Delta\Theta$.  Denoting the grid points by $\Theta$, initialize all elements of the following arrays to zero:
        \begin{enumerate}
            \item $p^+_A(\Theta)$ (resp.\ $p^+_B(\Theta)$), which accumulates the weight of trajectories entering state $A$ (resp.\ $B$); 
             \item $p^-_A(\Theta)$ (resp.\ $p^-_B(\Theta)$), which accumulates the weight of trajectories exiting state $A$ (resp.\ $B$); 
             
             \item $v_{AB}(\Theta)$ (resp.\ $v_{BA}(\Theta)$), which accumulates the weighted CV increments of trajectories exiting state $A$ and entering state $B$ (resp.\ exiting state $B$ and entering state $A$).
        \end{enumerate}
    \end{enumerate}
    }
    \item{Run NEUS essentially as described in Section 3.3 of ref.\ \onlinecite{dinner2018trajectory}.  Namely, at each iteration $l$: \label{neussteps}
    \begin{enumerate}
        \item Initialize all elements of a matrix $g^{(l)}$ with the same dimensions as $G$ (i.e., $2K\times 2K$) to zero.  \label{iterationstart}
        
        \item  For each stratum $j$, draw $N$ walkers from $\pi_j$. To be precise, for each walker, select an element of $\gamma_{ij}(dx)$ with probability $z_i G_{ij}$ and use the saved configuration to initialize the walker; this procedure represents \eqref{pi_j_expr2}.  Associate with the walker a pointer to the element of $\gamma_{ij}(dx)$.\label{drawwalkers}
        
        \item  Evolve each walker by unbiased dynamics until it  exits its initial stratum ($j$ in step \ref{drawwalkers}). \label{walkerexit}
 
        \item When a walker exits from stratum $j$ to stratum $k$, save its final configuration and associated pointer to $\gamma_{jk}(dx)$ and add 1 to $g^{(l)}_{jk}$.  Also save the sequence of $\theta$ values visited by the walker trajectory. \label{cvsave}
    
        \item Compute $G_{ij}$ as $$G_{ij}=\frac{\sum_{l'=L}^l g^{(l')}_{ij}}{(l-L+1)N},$$ where $L=\max[1,\min(l-L',L')]$, and $L'$ is chosen to minimize bias from the preliminary trajectories (i.e., to provide a burn-in period for the simulation).
    
        \item Update $z$ by solving \eqref{affineeig3}.% and $\pi_i$ using \eqref{pi_j_expr2}.  
        
        \item Check convergence criteria and go to step \ref{iterationstart} if not satisfied.  We examine both the change in the $z$ vector and the number of complete reactive trajectories.
        
    \end{enumerate}
    }
    \item When a walker enters $A\cup B$, reconstruct the sequence of walkers leading to that event from the last exit of $A\cup B$ and in turn the sequence of $\theta$ values and strata visited. For each stratum $i$ and each point $\theta(t)$ in the sequence, determine the nearest grid point $\Theta$ and  \label{tptptr}\label{tptsave}
    \begin{enumerate}
        \item \edits{increment $p_A^+(\Theta)$ (resp.\ $p_B^+(\Theta)$) by $z_i$} if the sequence \edits{terminated in} $A$ (resp.\ $B$);
        \item \edits{increment $p_A^-(\Theta)$ (resp.\ $p_B^-(\Theta)$) by  $z_i$} if the sequence \edits{originated in} $A$ (resp.\ $B$);
        \item  \edits{increment $v_{AB}(\Theta)$ (resp.\ $v_{BA}$) by $z_i[\theta(t+\tau)-\theta(t-\tau)]$} if the sequence \edits{originated in} $A$ and \edits{terminated in} $B$ (resp.\ \edits{originated in} $B$ and \edits{terminated in} $A$).
    \end{enumerate}
    \edits{Above, $z_i$ accounts for the weight of each trajectory segment, which is necessary to correct for the enhanced sampling.}

    \item{Construct the TPT quantities for reactive trajectories from $A$ to $B$ as follows (exchange $A$ and $B$ below for reactive trajectories from $B$ to $A$).
    \begin{enumerate}
        %\item When a walker enters $B$, reconstruct the sequence of walkers leading to that event \edits{from $(A\cup B)$} and in turn the sequence of $\theta$ values and strata visited. \label{tptptr}
        
        %\item For each stratum $i$ and each point $\theta(t)$ in the sequence, determine the nearest grid point $\Theta$ and add $z_i$ to $p^+_B(\Theta)$; if the sequence of $\theta$ values began in $A$, add $z_i[\theta(t+\tau)-\theta(t-\tau)]$ to $v_{AB}(\Theta)$ \edits{and add $z_i$ to $p^-_A(\Theta)$; if the sequence began in $B$, add $z_i$ to $p^-_B(\Theta)$}. \label{tptsave}
    
        \item Compute the forward committor to $B$ as $$q_+^\theta(\Theta)=\frac{p^+_B(\Theta)}{p^+_A(\Theta)+p^+_B(\Theta)}.$$
        %\edits{The denominator is the total weight of trajectories entering $B$.}
                \item Compute the backward committor from $A$ as $$q_-^\theta(\Theta)=\frac{p^-_A(\Theta)}{p^+_A(\Theta)+p^+_B(\Theta)}.$$
        Note that $p_A^-+p_B^-=p_A^++p_B^+$ is the total weight associated with all trajectories that start and end in $A\cup B$. 
        \item Compute the current from $A$ to $B$ as $$\edits{I}_{AB}^\theta (\Theta)=\frac{v_{AB}(\Theta)}{2\tau\Delta\Theta\sum_{\Theta'} [ p^+_A(\Theta') + p^+_B(\Theta')]}.$$
    \end{enumerate}
    }
        
\end{enumerate}
Figure \ref{fig:tptschematic} shows how the algorithm samples the ensemble of reactive trajectories in a piecewise fashion and how they are reconstructed.

%where $E_A$ is the distribution of trajectories that was last in $A$.
\section{Numerical Results}\label{numerical}

We demonstrate our method by analyzing the C$_\text{7ax}$ $\to$ C$_\text{7eq}$ transition of the alanine dipeptide ({\it N}-acetyl-alanyl-{\it N'}-methylamide) in vacuum.  The peptide is represented by the CHARMM36m force field  \cite{MacKerell1998, Best2012, Huang2017}.  The simulations are performed at 300 K with the SD (stochastic dynamics) integrator\cite{goga2012efficient} in GROMACS 5.1.4 \cite{Abraham2015}; LINCS was used to constrain the lengths of bonds to hydrogen atoms; the step size was 1 fs, and the friction coefficient was $10 \text{ ps}^{-1}$. We used PLUMED 2.3 \cite{Bonomi2009, Tribello2014, PLUMED2019} to extract collective variables; simulations were terminated when walkers left their strata using the PLUMED COMMITTOR function.

The metastable states and strata are defined in terms of the $\phi$ and $\psi$ dihedral angles. The metastable states are taken to be circles of radius $20^\circ$ around  $(\phi, \psi) = (-83^\circ,75^\circ)$ and $(70^\circ,-70^\circ)$, which correspond to the minima of the potential of mean force determined below. 
The system was initialized from preliminary trajectories that were generated by using Steered Transition Path Sampling (STePS) \cite{guttenberg2012steered} to drive the system from each metastable state to the other.  STePS builds up trajectories by repeatedly shooting bursts of short segments and then preferentially selecting those that make forward progress for continuation.  Trajectories were initialized from points in the equilibrium ensemble within each metastable state, and progress was measured by the distance to the center of the other metastable state.  The segment length was 1 ps, and the bias threshold was 0.75 (i.e., forward trajectories are selected with a probability of 0.75 or their probability estimated from the burst, whichever is higher).  Initially, 10 trajectories were released in each burst; subsequently, we varied the number with the progress distance such as to expect two forward segments in each burst based on statistics of the previous run.  We do not use the statistics to correct for the STePS bias---i.e., each reactive trajectory is weighted equally.

\begin{figure}[h!]
\centering
\includegraphics[scale=0.3]{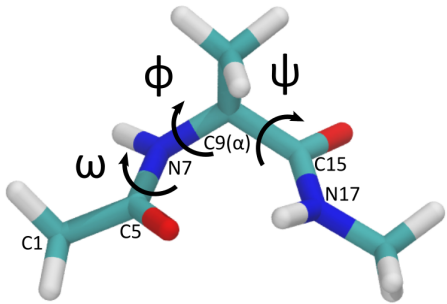}
\caption{Alanine dipeptide with the three dihedral angles labeled:  $\phi$ (C5-N7-C9($\alpha$)-C15), $\psi$ (N7-C9($\alpha$)-C15-N17), and (C1-C5-N7-C9($\alpha$)). Colors are cyan for carbon, blue for nitrogen, white for hydrogen, and red for oxygen.\label{fig:angles}
}
\end{figure}

The strata form a partition of unity; their centers are at $(-7+2n)30^\circ$ for integer $n\in[1,6]$, and their widths are $100^\circ$, so they overlap $40^\circ$ both ways, and the strata wrap around to enforce periodicity. We release $10$ walkers in every stratum in each iteration and run each iteration until all 10 walkers exit their stratum.  The collective variables $\phi$ and $\psi$ and their increments are saved every 100 fs. The lag time used to compute the increments ($\tau$ in \eqref{JABtheta}) was 2 ps.  The simulation is run until we see at least $1000$ crossings in each direction between the metastable states, i.e., walkers that cross from $j\leq K$ to $j>K$ and the other way. We used a burn-in period of $L'=100$ iterations. The full simulation was %$139,834,245$ 
approximately 140 million steps.
Though we stratify in $\phi$ and $\psi$, we can project the results onto other variables as well, and we also consider $\omega$ (Figure \ref{fig:angles}). Statistics are accumulated for 50 values of each dihedral angle.

\begin{figure}[bt]
\centering
\includegraphics[scale=0.7]{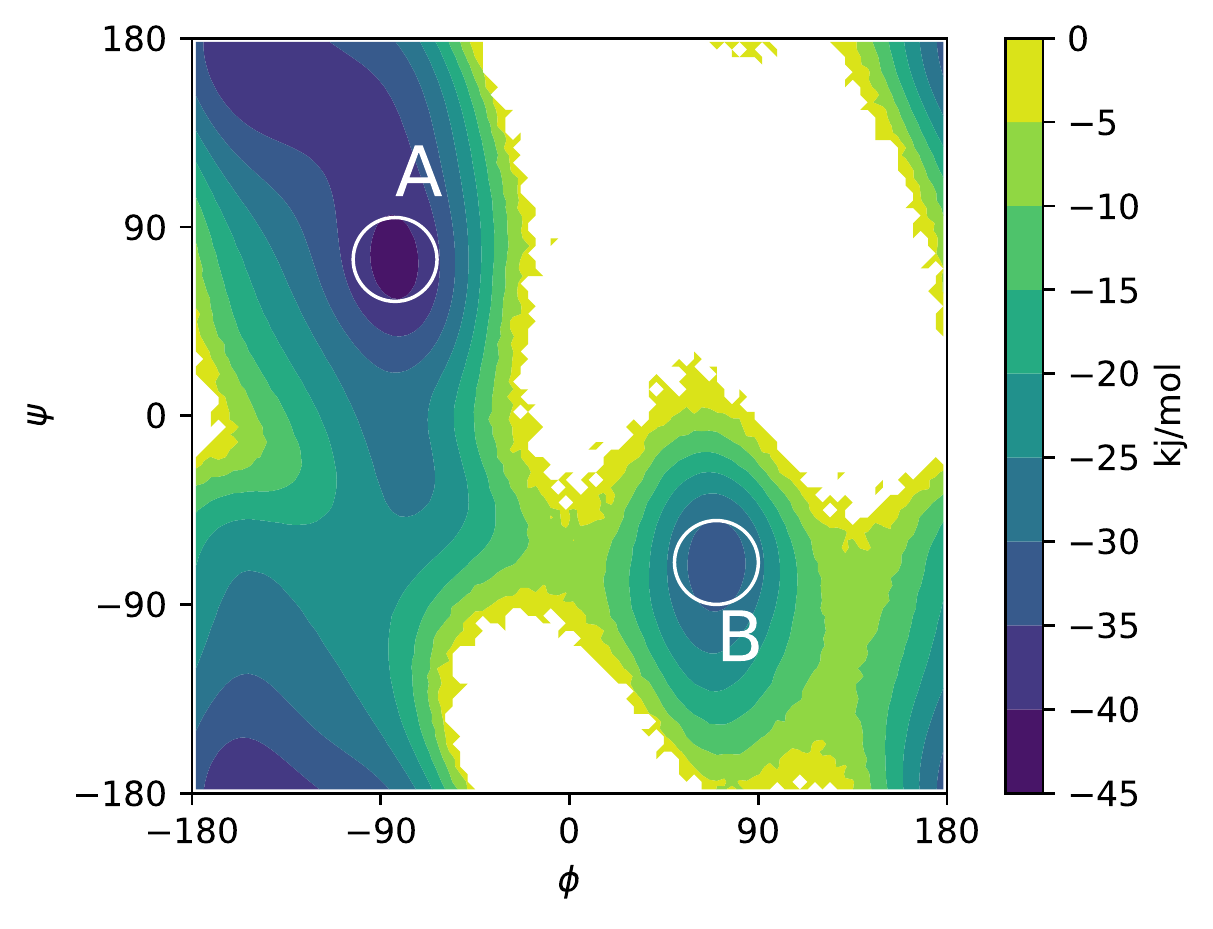}
\includegraphics[scale=0.7]{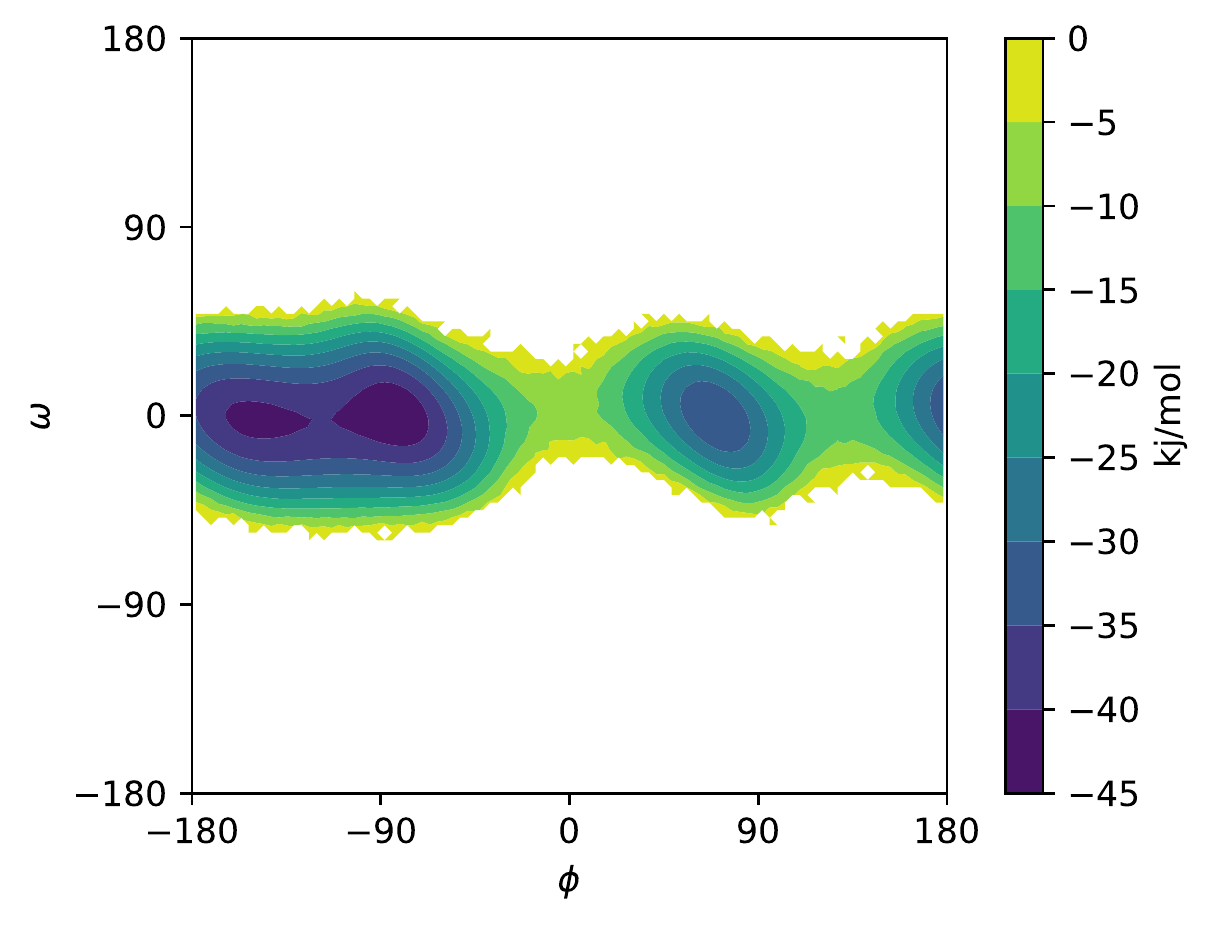}
\caption{Potentials of mean force computed from the NEUS simulations:  projections on (top) $\phi$ and $\psi$ and (bottom) $\phi$ and $\omega$. The stable states are marked with white circles on the top plot.\label{PMF_adp}
}
\end{figure}

%(so I actually used STePS to initialize for reasons that are now irrelevant. Essentially just need to initialize with points in a basin (convenient to have points in both basins, but not necessary). Since I'd already run steps, I kept using those as my initial points) 

%Initialization is also non-trivial for this method, since walkers are indexed according to their history, and structures chosen from an equilibrium distribution will not contain that information. Instead, one needs to seed with reactive trajectories.  Generally these cannot be obtained from direct simulations.  One option is to run NEUS with a third index and use it to feed the flux lists of the pre-defined indices, but this is also a computationally wasteful process. 

\begin{figure}[!]
%\centering
\includegraphics[scale=0.7]{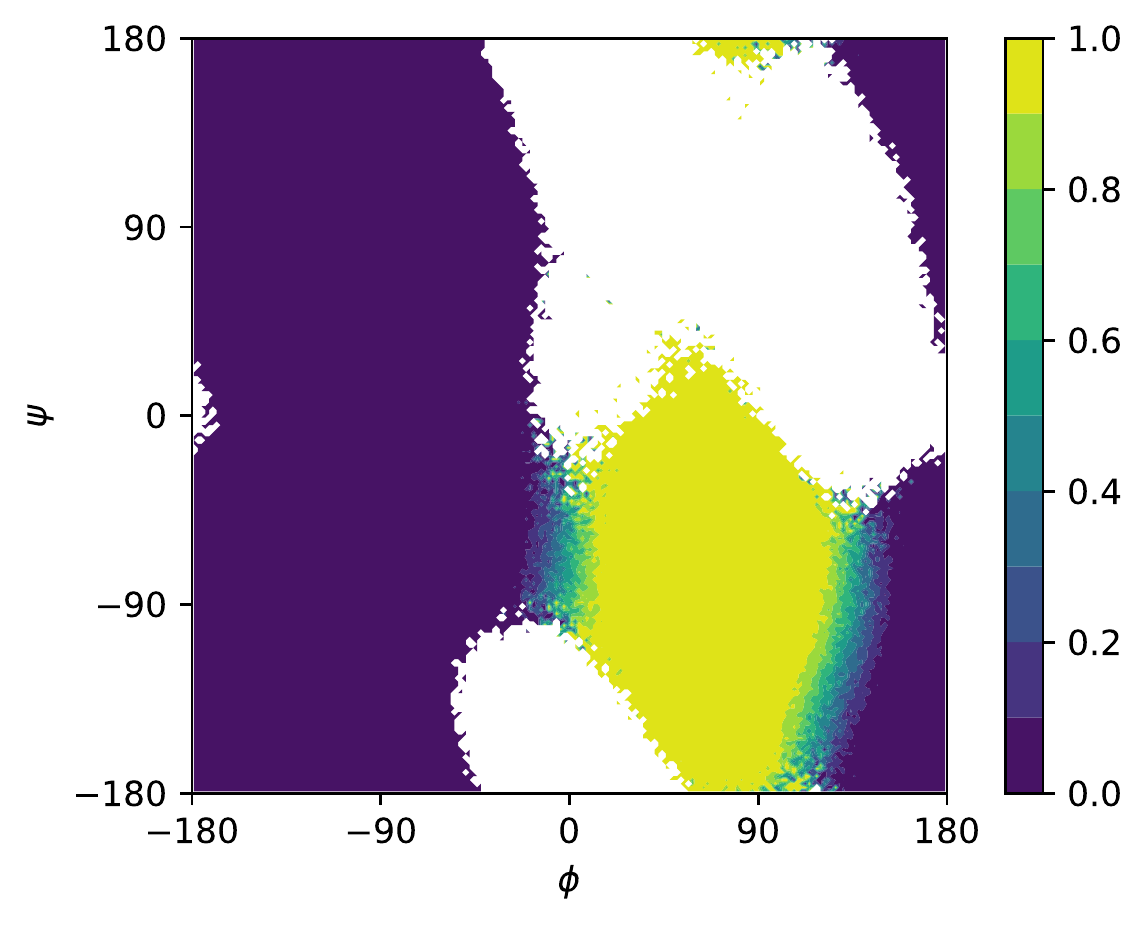}
\includegraphics[scale=0.7]{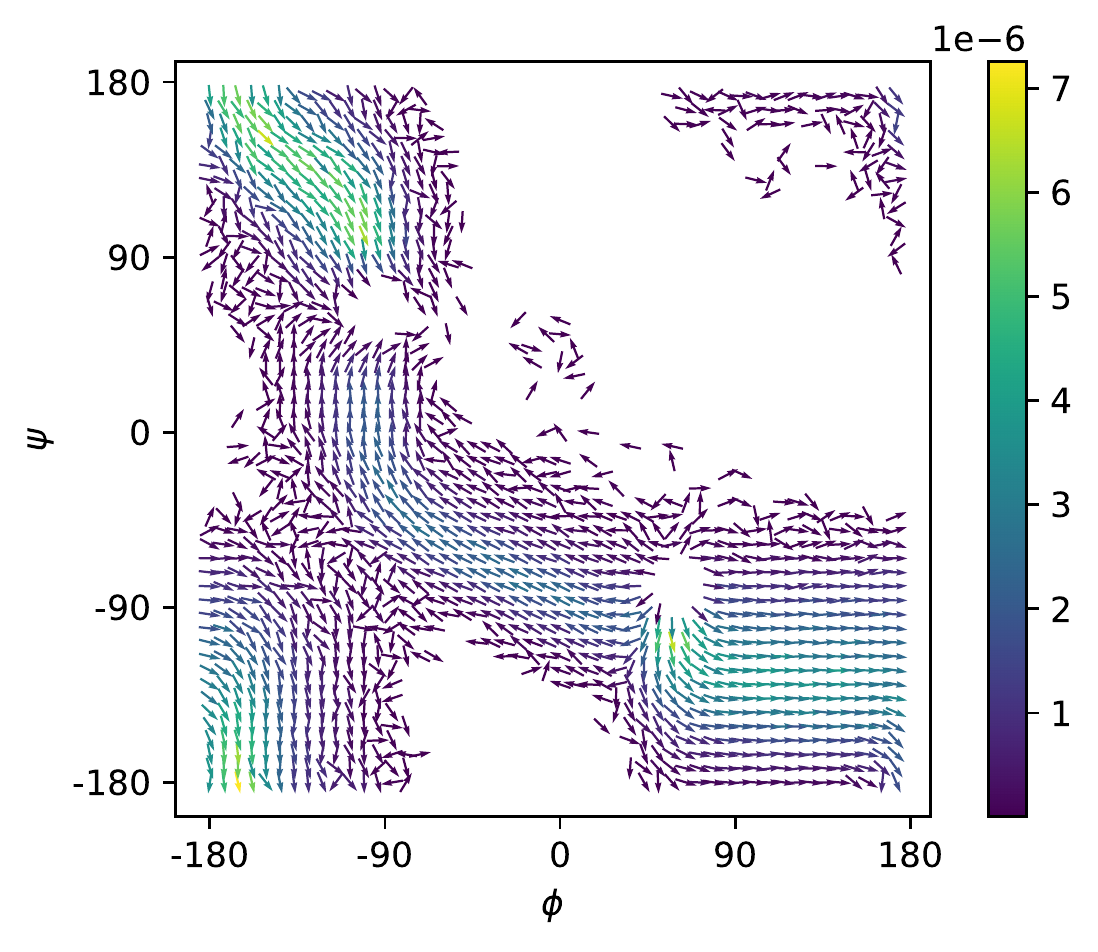}
\includegraphics[scale=0.7]{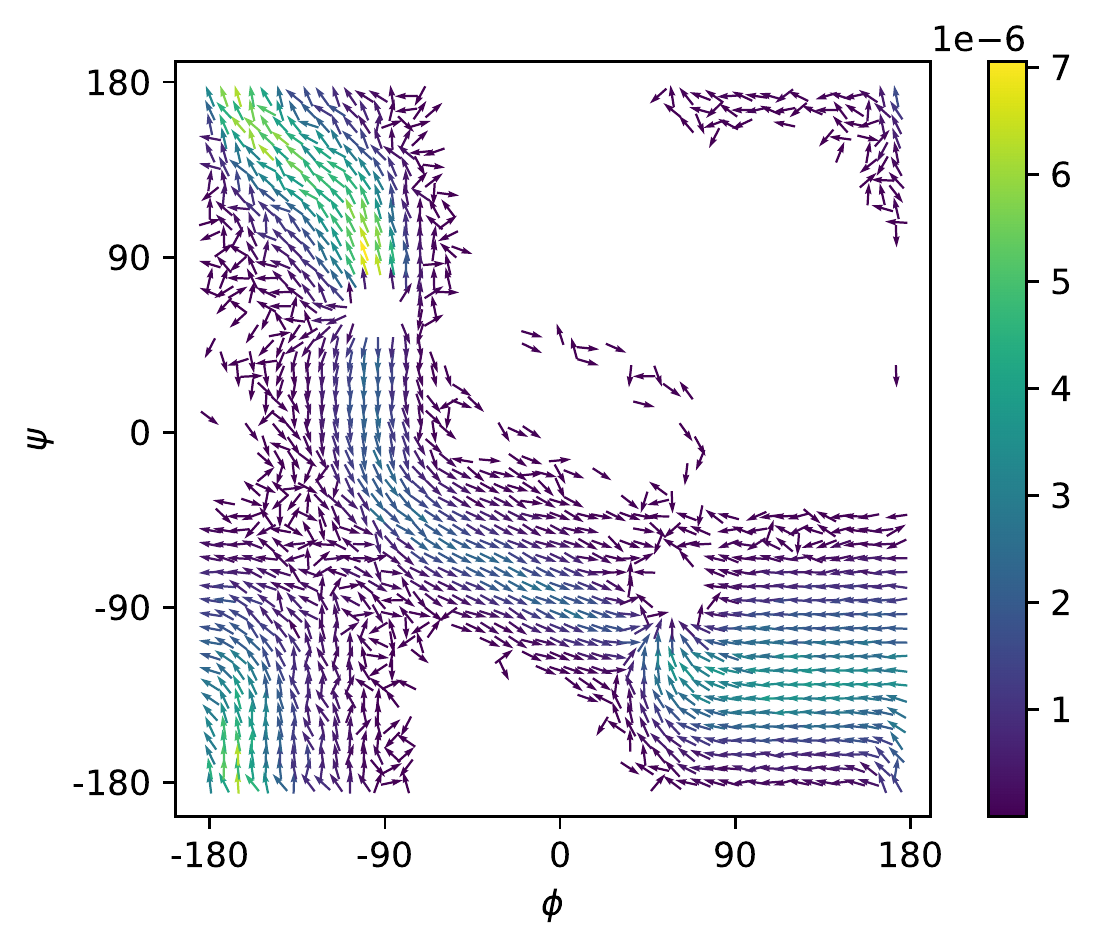}
\caption{Committor (top) and forward (middle) and backward (bottom) reactive currents for the C7ax $\to$ C7eq transition of the alanine dipeptide, plotted on the $\phi\psi$-plane. \label{fig:phipsiproj}
}
\end{figure}

\begin{figure}[h!]
\centering
\includegraphics[scale=0.7]{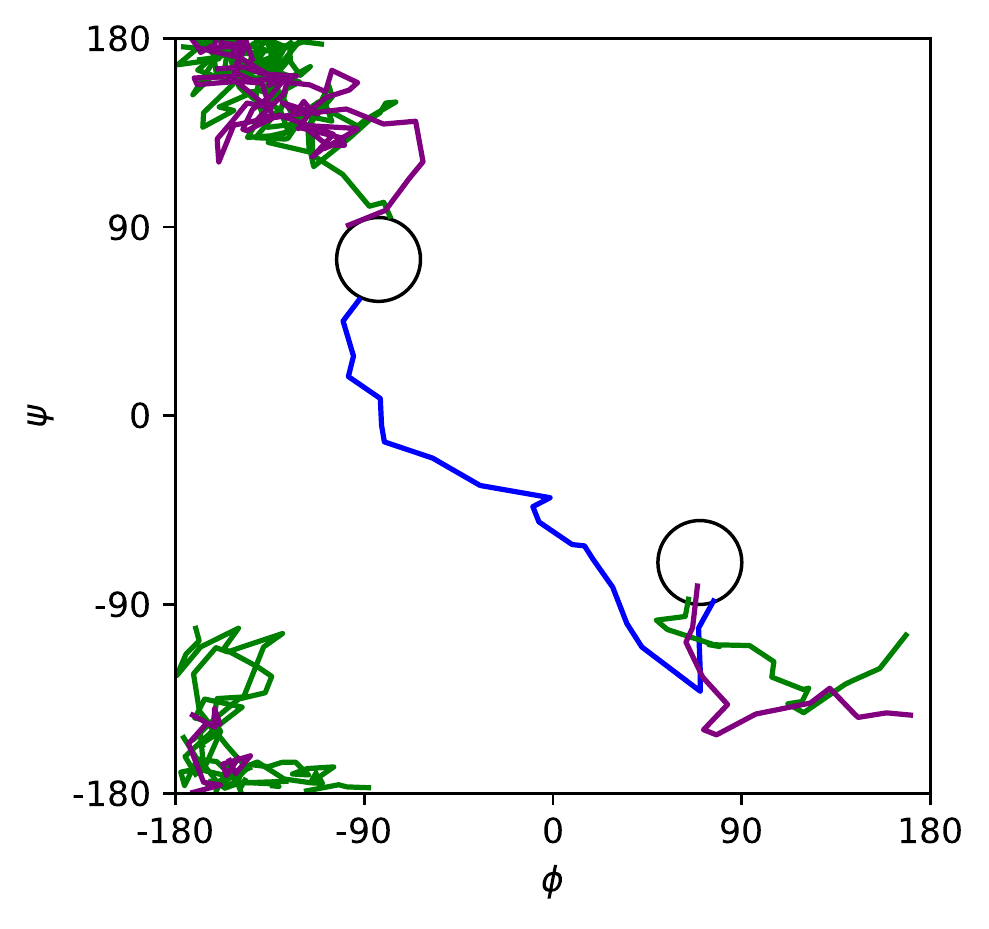}
\caption{ Example trajectories plotted on the $\phi\psi$-plane. Dots mark the centers of the metastable states.
\label{fig:examples}
}
\end{figure}

\begin{figure}[!]
\centering
\includegraphics[scale=0.7]{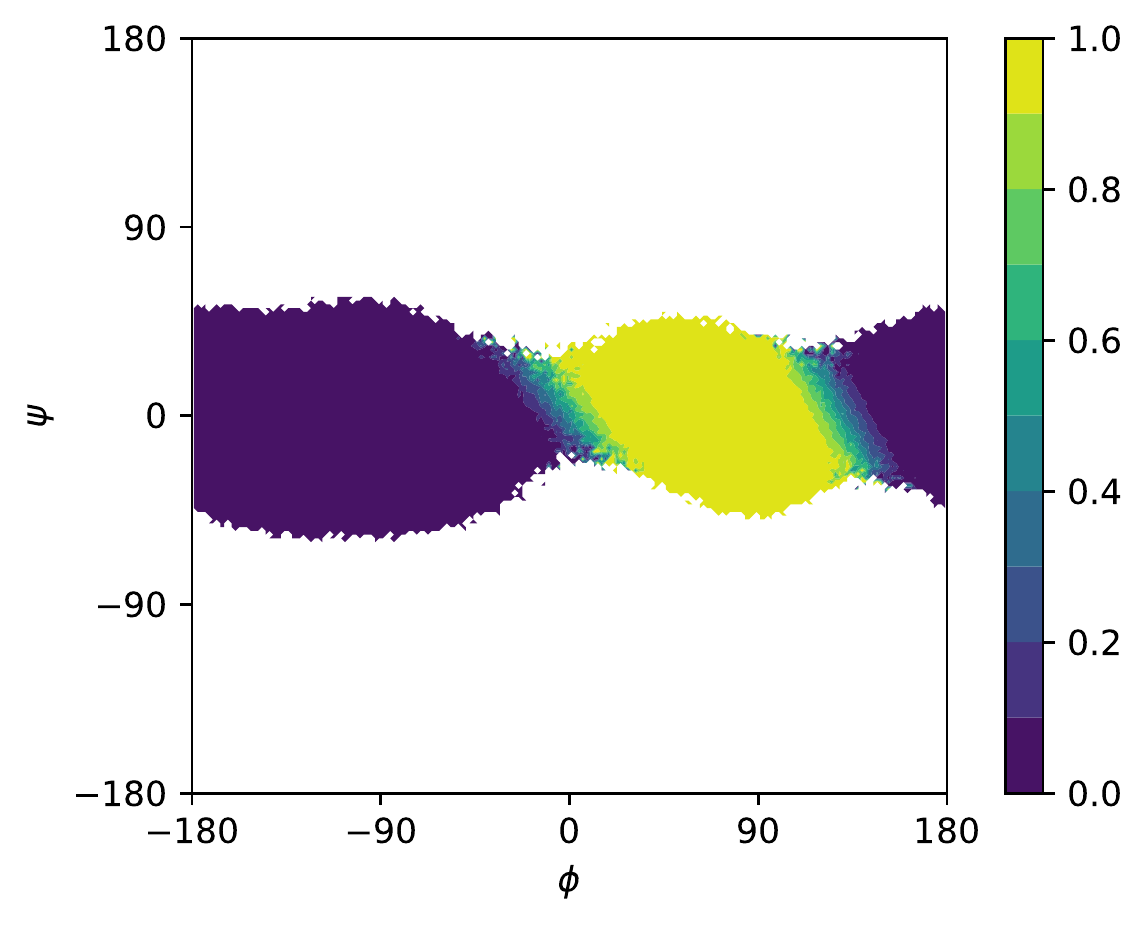}
\includegraphics[scale=0.7]{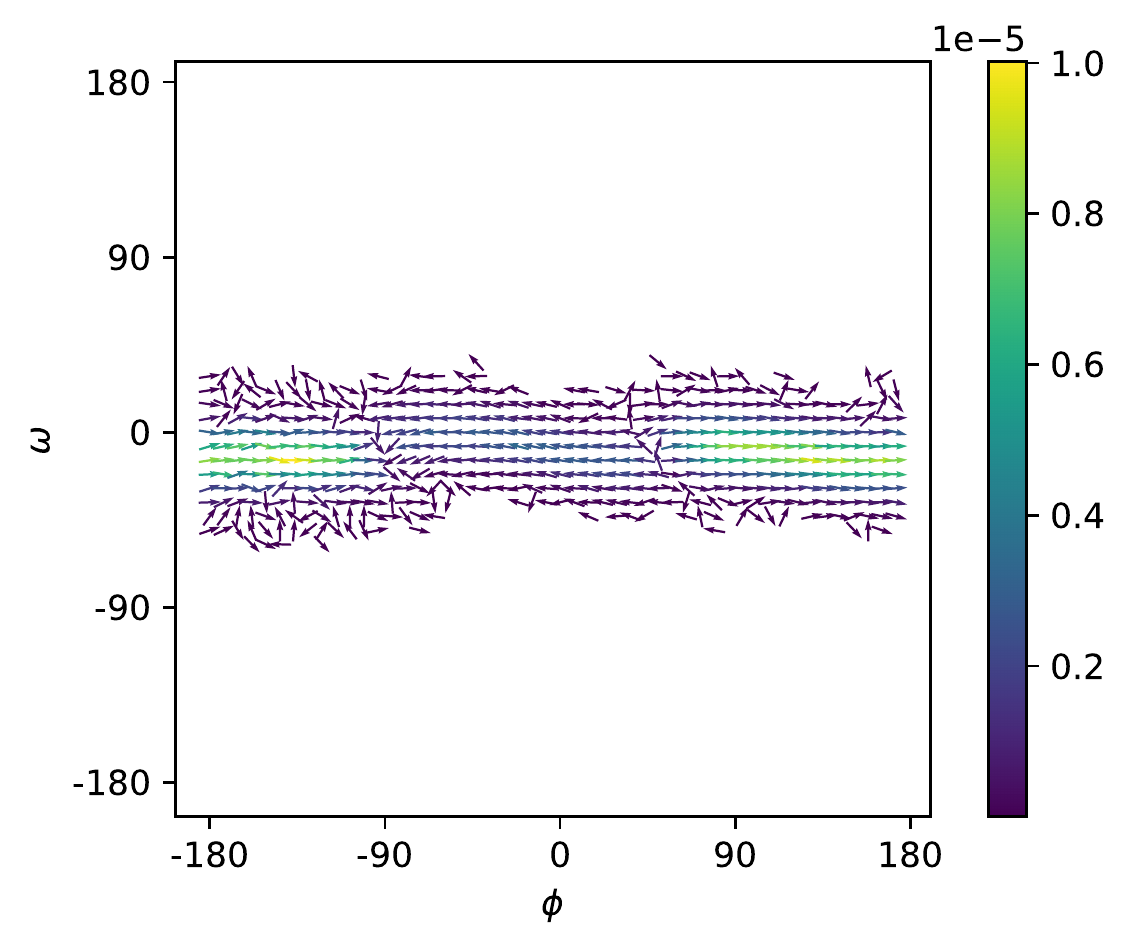}
\includegraphics[scale=0.7]{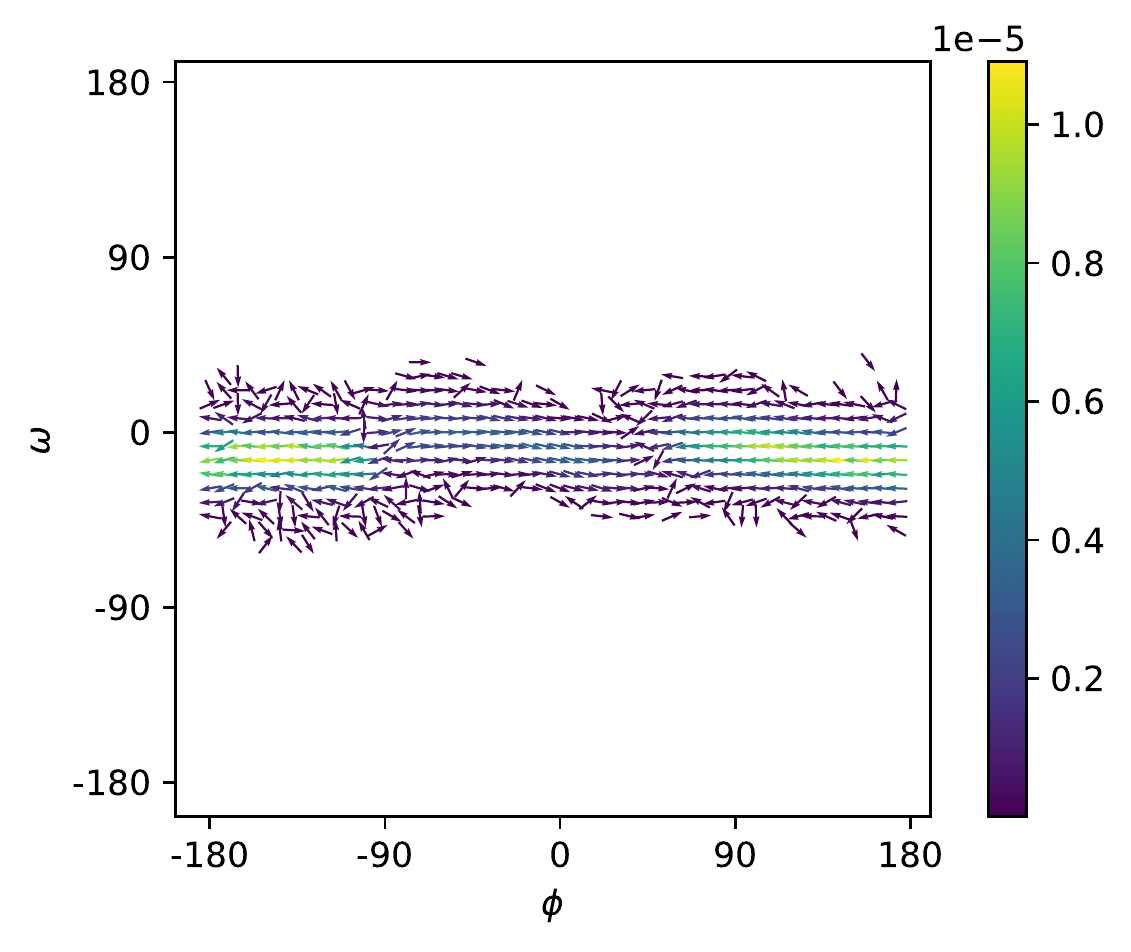}
\caption{Committor (top) and forward (middle) and backward (bottom) reactive currents for the C7ax $\to$ C7eq transition of the alanine dipeptide, plotted on the $\phi\omega$-plane.  For the reactive currents, only grid points with data from at least 5 trajectories are shown. \label{fig:phiomegaproj}
}
\end{figure}

Potentials of mean force (PMF) computed by NEUS are shown in Figure \ref{PMF_adp}.  
In Figure \ref{fig:phipsiproj} we show the forward committor and the $A\to B$ and $B\to A$ currents in the $\phi\psi$-plane.  We can see the clear existence of two main pathways, one of which cuts diagonally across the middle of the figure and one of which cuts across the upper left and lower right corners.  Three representative trajectories consistent with the pathways are shown in Figure \ref{fig:examples}. 
In Figure \ref{fig:phiomegaproj}, we show the commitor and currents in the $\phi\omega$-plane. We observe a diagonal structure in these plots that is consistent with coupling between distortion of the peptide plane and $\phi$ noted previously \cite{bolhuis2000reaction}.

To assess the accuracy of our committor estimates, we compute the committor as a function of all three dihedral angles (not shown), select configurations predicted to have $0.49<q_+^\theta<0.51$, and evaluate their committors by shooting 20 independent simulations from each configuration.  The resulting histogram of values is peaked at $q_+\approx 0.5$, as desired.   
To characterize the convergence of the committor estimates, we compute the backwards committor and plot $q_+^\theta + q_-^\theta -1$, which should be close to zero for this system (Figures 9 and 10); the deviation provides an indication of the numerical error. We see that after 250 crossings between the metastable states the error overall is small, with the largest deviations in regions with high free energy close to where $q_+^\theta=0.5$.  These deviations decrease with additional crossings. 

\begin{figure}[h!]
\centering
\includegraphics[scale=0.5]{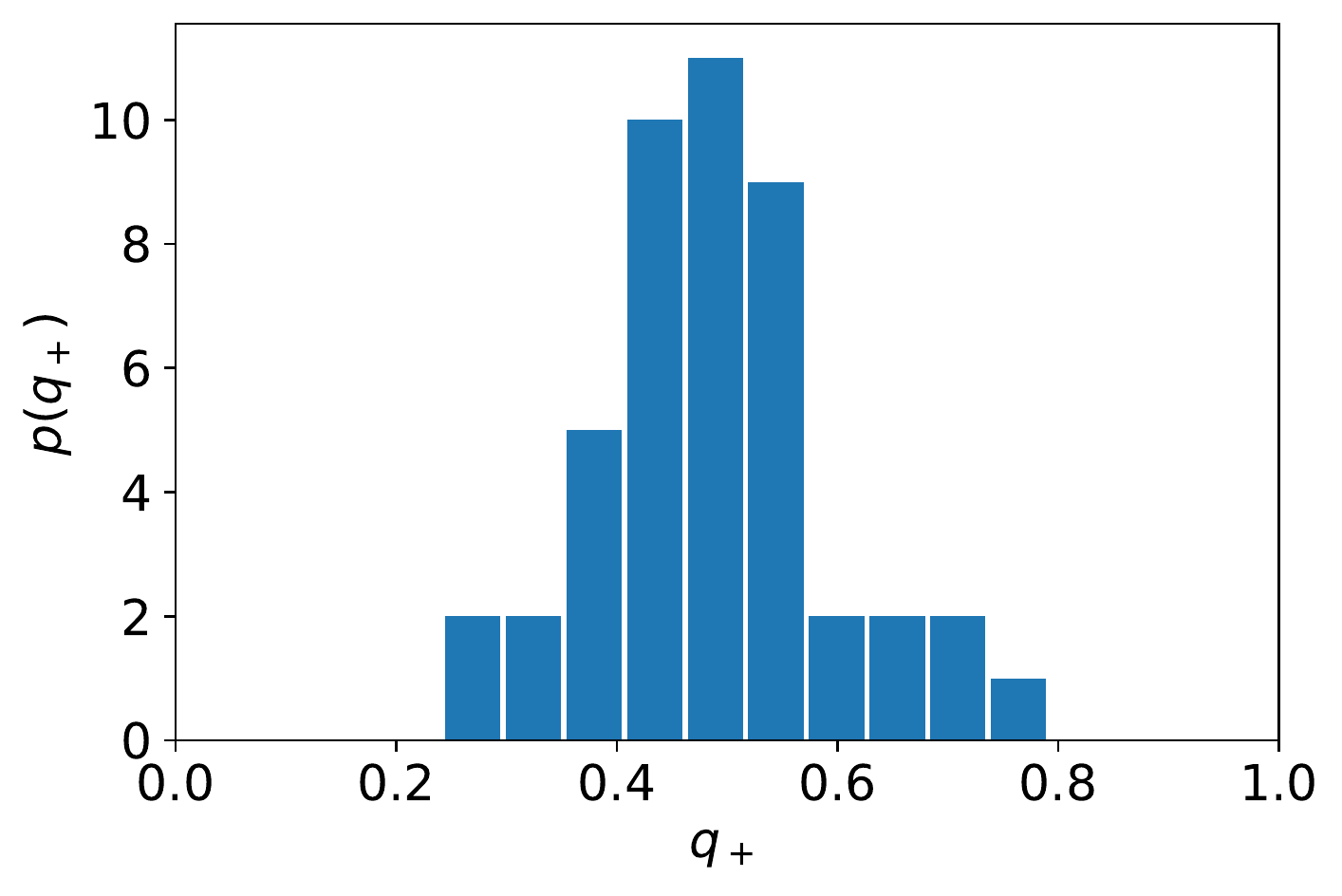}
\caption{Histogram of committors computed by shooting for %probability of hitting basin $A$ 
137 structures predicted to have $0.49<q_+^\theta<0.51$ for reaction from $A$ to $B$.  20 independent simulations were used for each configuration.
\label{fig:separatrix}
}
\end{figure}

To assess the reactive current estimates, we compute the rate as a function of the number of crossings two ways.  First we directly sum the flux into $B$\cite{vanden2009exact,dickson2009separating}:
\begin{equation}\label{neusrate}
    R_{\rm NEUS}=\frac{1}{\sum_{j=1}^{2K} z_j}\sum_{j\leq K} z_j \frac{n^B_j}{T_j}  ,
\end{equation}
where $n^B_j$ is the number of reactive trajectories that enter $B$ from stratum $j$, $T_j$ is the total time simulated in that stratum, and the sum over $j\leq K$ selects for contributions from trajectories that were last in $A$. 
Second, we sum the reactive currents that cross a dividing surface $S$:
\begin{equation}\label{tptrate}
    R_{\rm TPT}=\sum_S \edits{I}_{AB}^\theta (\Theta) \cdot \hat{n}_S \Delta_S,
\end{equation}
where $\hat{n}_S$ is the unit vector normal to $S$, and $\Delta_S$ is a differential element with the same dimension as $S$. Since the collective variable space is periodic, there are two pathways, and we average over a set of surfaces for each.  Specifically, we use vertical lines in the $\phi\psi$ plots:  (i) $\phi=(-57.6+7.2n)^\circ$ for integer $n$ in $[0, 15]$, which cut through the middle of Figure \ref{fig:phipsiproj}, and (ii) $\phi=(-172.8+7.2n)^\circ$ for integer $n$ in $[0, 9]$ and $[37, 49]$, which line the left and right sides of Figure \ref{fig:phipsiproj}.

\begin{figure}[!]
\centering
\includegraphics[scale=0.5]{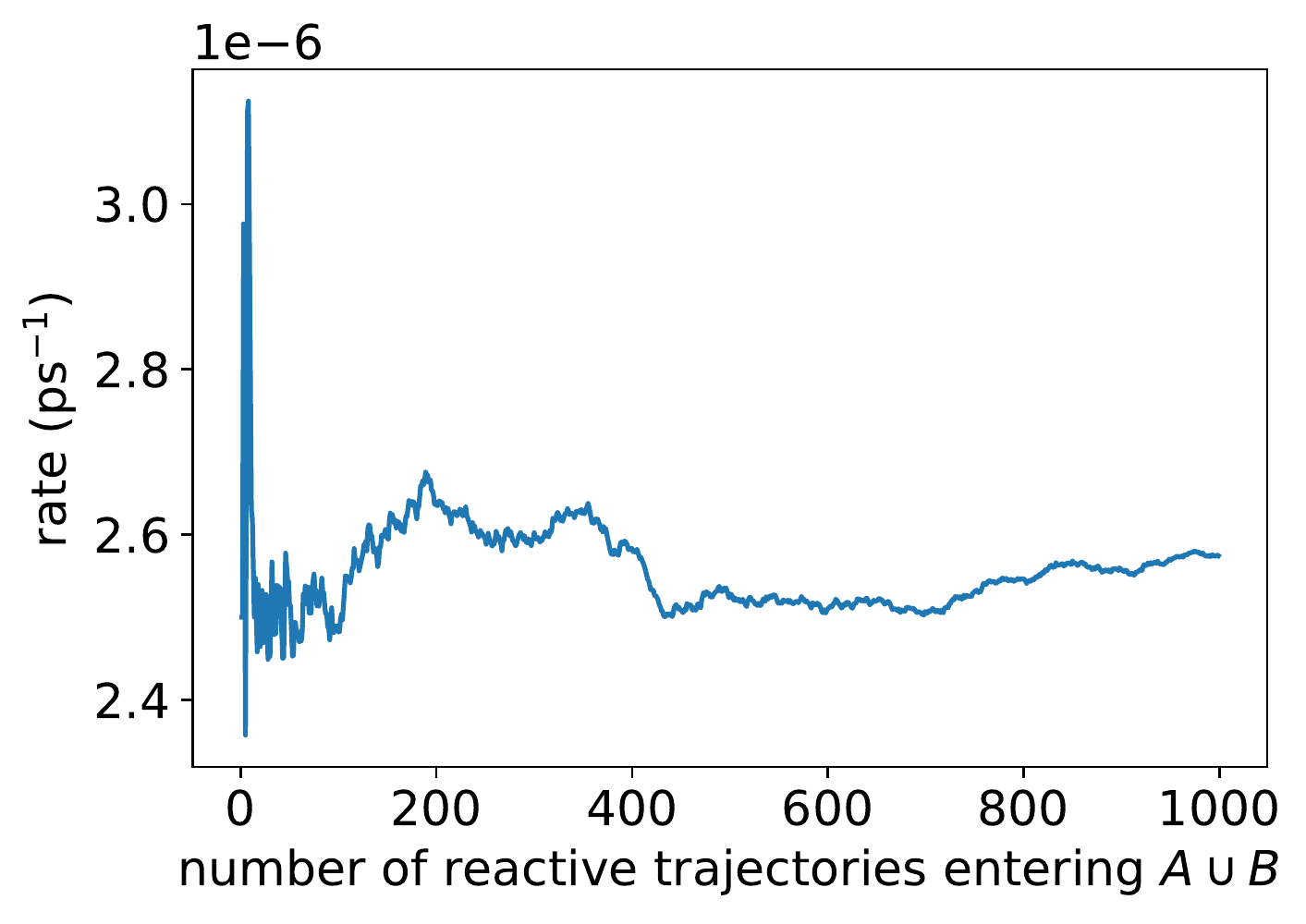}
\includegraphics[scale=0.5]{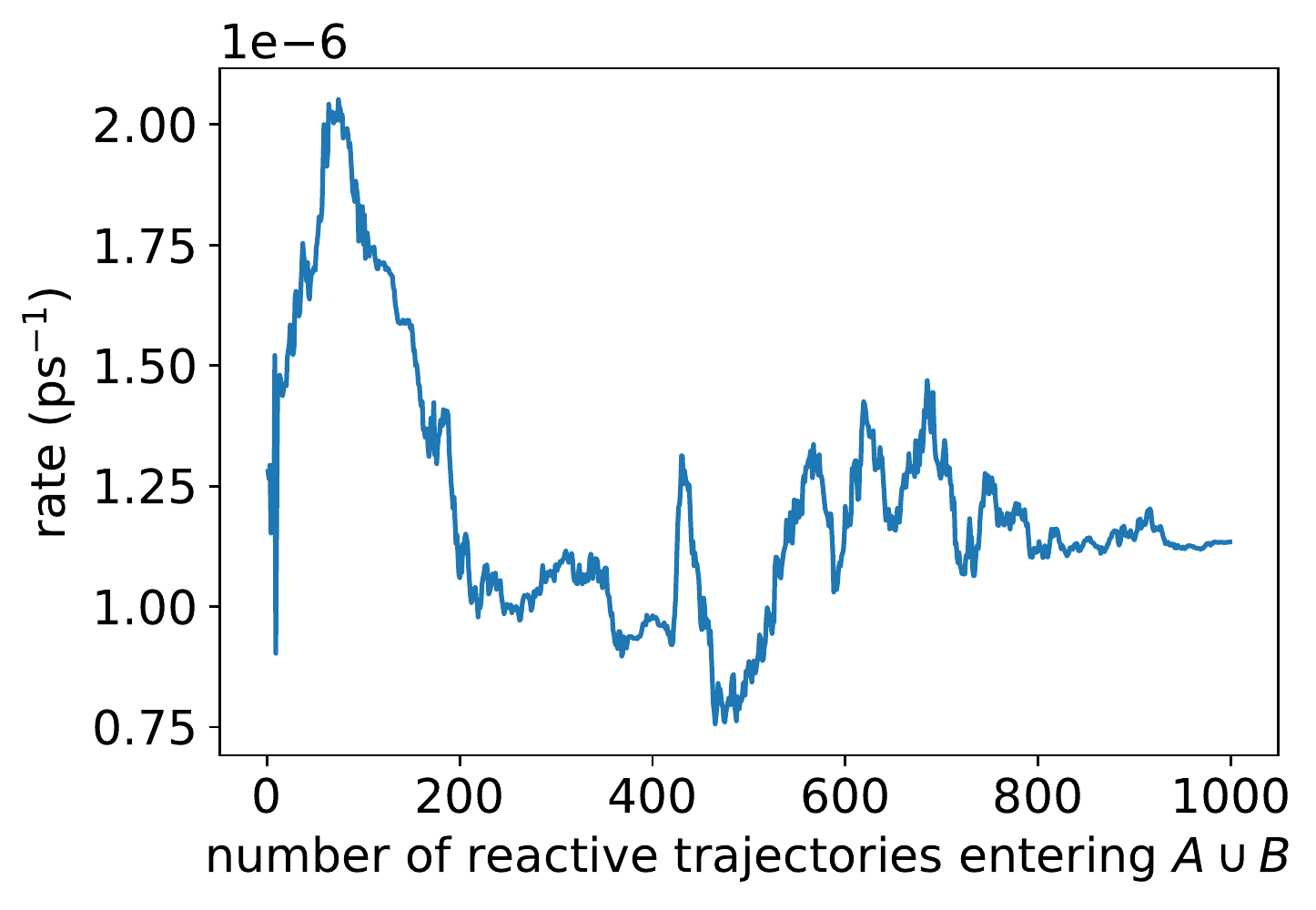}
\caption{
Rates computed directly from the flux into $B$ (top) and by integrating the  reactive currents (bottom).  An equal number of crossings between metastable states each way is used for each point on the horizontal axis.
\label{fig:rates}
}
\end{figure}

The two estimates are about a factor of two different from each other (Figure \ref{fig:rates}); we also obtain a direct estimate of $1.4 \times 10^{-6}$ ps$^{-1}$ from an unbiased trajectory of length 2.5 $\mu$s.  We consider the three estimates to be good agreement given that the rate can be a very challenging statistic to converge to even its order of magnitude.  We expect \eqref{neusrate} to be most precise (as evidenced by the scale of fluctuations in Figure \ref{fig:rates}) and recommend its use; we consider \eqref{tptrate} only as a means of validating the reactive currents.
That said, we can use the currents to obtain the fluxes associated with each of the pathways.  We find them to be $R_{\rm TPT}=4.6 \times 10^{-7}$ ps$^{-1}$ for the pathway that crosses $\phi=0^\circ$ and $R_{\rm TPT}=8.1 \times 10^{-7}$ ps$^{-1}$ for the pathway that crosses $\phi=180^\circ$.

To further assess the reactive currents, we also compute $\edits{I}_{AB}^\theta(\Theta)+J_{BA}^\theta(\Theta)$ (Figure 11).  We expect this sum to be close to but not exactly equal to zero given that the integrator is underdamped.  We see that this is the case, with the only significant deviations close to the metastable states.  Put together, these results validate the method.

%In Figure \ref{fig:sketch}, we suggest the two main pathways that are representative of this transition, with the solid line being the more probable path and the dotted line being the less probable path. Note that these are just sketched by hand and are not mathematically representative of free energy minimum paths.

%\begin{figure}[h!]
%\centering
%\includegraphics[scale=0.7]{sketch.pdf}
%\caption{ Sketch of putative reaction tube centers overlayed on reactive current in the $\phi\psi$-plane.  \label{fig:sketch}
%}
%\end{figure}

\section{Conclusions}

In this paper, we investigated the computation of committors and reactive currents with trajectory stratification.  The fundamental challenge is that these quantities require knowledge of the metastable states in which trajectories begin and end, but this information cannot be obtained directly from most walkers within the simulation.  To address this issue, we store the collective variable values visited by each walker and the associated collective variable increments sampled, as well as a pointer to the previous walker.  This enables us to reconstruct the sequence of walkers that gave rise to a trajectory once it reaches a metastable state, despite the fact that the number of trajectories to which a single walker can contribute grows exponentially with the number of walkers that follow it.  
The current work builds on previous trajectory stratification studies that showed that computation of the rate requires separating the ensemble of trajectories based on the metastable state in which they originate \cite{dickson2009separating,vanden2009exact}.  As in previous trajectory stratification studies, the results are exact in the limit of infinite sampling, in contrast to MSMs and DGA. 

Although we demonstrated the approach with NEUS, it is general and can be applied to other algorithms that sample trajectory segments \cite{huber_weighted-ensemble_1996,zhang2007efficient,zuckerman2017weighted,bolhuis2003transition,van2003novel,allen2005sampling,allen2009forward,bello2015exact}.
By the same token, we focused on a simple numerical example that permits validation of the results. Extending the approach to more complex systems should be straightforward, so long as collective variables that enable an efficient stratification and the accumulation of statistics can be identified.  This challenge is system specific, though methods for facilitating the selection of collective variables based on simulation data exist (e.g., refs.\ \onlinecite{ma_automatic_2005,strahan2021long}).  In the case of NEUS, the most significant challenge may be that the lists representing $\gamma_{ij}(dx)$ can become large (up to 5500 elements in the present example).  \edits{Although this issue has not been limiting for complex systems treated to date \cite{dickson2011flow}, it} is an important practical consideration that we leave for future work.

%First, we derive an expression for the reactive current in collective variable space that can be written as an expectation over an infinitely long trajectory, and cast it as an observable that can be accumulated by reactive trajectories. We present a trajectory stratification approach to computing these quantities. However, note that the method presented casts the trajectory dependent variables as observables carried forward to the flux into basins, so it can be adapted to any path sampling scheme, and is independent of stratification scheme.

%We demonstrate the trajectory stratified approach to computing tpt quantities on a transition in alanine dipeptide in vaccuum, and get a well converged reactive current that illuminates two main transition tubes for the transition.

\section{Supplementary Material}

Justification of \eqref{JABtheta}, committor convergence plots, sums of forward and backward reactive currents.

\section{Acknowledgments}

The authors wish to thank Chatipat Lorpaiboon and John Strahan for useful discussions and critical readings of the manuscript.  This work was supported by National Institutes of Health award R35 GM136381 and National Science Foundation award DMS-2054306.

%\bibliographystyle{unsrt}
%\bibliography{tpt,trp-cage}

\begin{thebibliography}{50}%
\makeatletter
\providecommand \@ifxundefined [1]{%
 \@ifx{#1\undefined}
}%
\providecommand \@ifnum [1]{%
 \ifnum #1\expandafter \@firstoftwo
 \else \expandafter \@secondoftwo
 \fi
}%
\providecommand \@ifx [1]{%
 \ifx #1\expandafter \@firstoftwo
 \else \expandafter \@secondoftwo
 \fi
}%
\providecommand \natexlab [1]{#1}%
\providecommand \enquote  [1]{``#1''}%
\providecommand \bibnamefont  [1]{#1}%
\providecommand \bibfnamefont [1]{#1}%
\providecommand \citenamefont [1]{#1}%
\providecommand \href@noop [0]{\@secondoftwo}%
\providecommand \href [0]{\begingroup \@sanitize@url \@href}%
\providecommand \@href[1]{\@@startlink{#1}\@@href}%
\providecommand \@@href[1]{\endgroup#1\@@endlink}%
\providecommand \@sanitize@url [0]{\catcode `\\12\catcode `\$12\catcode
  `\&12\catcode `\#12\catcode `\^12\catcode `\_12\catcode `\%12\relax}%
\providecommand \@@startlink[1]{}%
\providecommand \@@endlink[0]{}%
\providecommand \url  [0]{\begingroup\@sanitize@url \@url }%
\providecommand \@url [1]{\endgroup\@href {#1}{\urlprefix }}%
\providecommand \urlprefix  [0]{URL }%
\providecommand \Eprint [0]{\href }%
\providecommand \doibase [0]{http://dx.doi.org/}%
\providecommand \selectlanguage [0]{\@gobble}%
\providecommand \bibinfo  [0]{\@secondoftwo}%
\providecommand \bibfield  [0]{\@secondoftwo}%
\providecommand \translation [1]{[#1]}%
\providecommand \BibitemOpen [0]{}%
\providecommand \bibitemStop [0]{}%
\providecommand \bibitemNoStop [0]{.\EOS\space}%
\providecommand \EOS [0]{\spacefactor3000\relax}%
\providecommand \BibitemShut  [1]{\csname bibitem#1\endcsname}%
\let\auto@bib@innerbib\@empty
%</preamble>
\bibitem [{\citenamefont {Ma}\ and\ \citenamefont
  {Dinner}(2005)}]{ma_automatic_2005}%
  \BibitemOpen
  \bibfield  {author} {\bibinfo {author} {\bibfnamefont {A.}~\bibnamefont
  {Ma}}\ and\ \bibinfo {author} {\bibfnamefont {A.~R.}\ \bibnamefont
  {Dinner}},\ }\bibfield  {title} {\enquote {\bibinfo {title} {Automatic method
  for identifying reaction coordinates in complex systems},}\ }\href {\doibase
  10.1021/jp045546c} {\bibfield  {journal} {\bibinfo  {journal} {Journal of
  Physical Chemistry B}\ }\textbf {\bibinfo {volume} {109}},\ \bibinfo {pages}
  {6769--6779} (\bibinfo {year} {2005})}\BibitemShut {NoStop}%
\bibitem [{\citenamefont {Vanden-Eijnden}(2006)}]{Vanden-Eijnden2006}%
  \BibitemOpen
  \bibfield  {author} {\bibinfo {author} {\bibfnamefont {E.}~\bibnamefont
  {Vanden-Eijnden}},\ }\enquote {\bibinfo {title} {Transition path theory},}\
  in\ \href {\doibase 10.1007/3-540-35273-2_13} {\emph {\bibinfo {booktitle}
  {Computer Simulations in Condensed Matter Systems: From Materials to Chemical
  Biology Volume 1}}},\ \bibinfo {editor} {edited by\ \bibinfo {editor}
  {\bibfnamefont {M.}~\bibnamefont {Ferrario}}, \bibinfo {editor}
  {\bibfnamefont {G.}~\bibnamefont {Ciccotti}}, \ and\ \bibinfo {editor}
  {\bibfnamefont {K.}~\bibnamefont {Binder}}}\ (\bibinfo  {publisher} {Springer
  Berlin Heidelberg},\ \bibinfo {address} {Berlin, Heidelberg},\ \bibinfo
  {year} {2006})\ pp.\ \bibinfo {pages} {453--493}\BibitemShut {NoStop}%
\bibitem [{\citenamefont {E.}\ and\ \citenamefont
  {Vanden-Eijnden}(2006)}]{e_towards_2006}%
  \BibitemOpen
  \bibfield  {author} {\bibinfo {author} {\bibfnamefont {W.}~\bibnamefont
  {E.}}\ and\ \bibinfo {author} {\bibfnamefont {E.}~\bibnamefont
  {Vanden-Eijnden}},\ }\bibfield  {title} {\enquote {\bibinfo {title} {Towards
  a theory of transition paths},}\ }\href {\doibase 10.1007/s10955-005-9003-9}
  {\bibfield  {journal} {\bibinfo  {journal} {Journal of Statistical Physics}\
  }\textbf {\bibinfo {volume} {123}},\ \bibinfo {pages} {503} (\bibinfo {year}
  {2006})}\BibitemShut {NoStop}%
\bibitem [{\citenamefont {Metzner}, \citenamefont {Schütte},\ and\
  \citenamefont {Vanden-Eijnden}(2006)}]{metzner_illustration_2006}%
  \BibitemOpen
  \bibfield  {author} {\bibinfo {author} {\bibfnamefont {P.}~\bibnamefont
  {Metzner}}, \bibinfo {author} {\bibfnamefont {C.}~\bibnamefont {Schütte}}, \
  and\ \bibinfo {author} {\bibfnamefont {E.}~\bibnamefont {Vanden-Eijnden}},\
  }\bibfield  {title} {\enquote {\bibinfo {title} {Illustration of transition
  path theory on a collection of simple examples},}\ }\href {\doibase
  10.1063/1.2335447} {\bibfield  {journal} {\bibinfo  {journal} {Journal of
  Chemical Physics}\ }\textbf {\bibinfo {volume} {125}},\ \bibinfo {pages}
  {084110} (\bibinfo {year} {2006})}\BibitemShut {NoStop}%
\bibitem [{\citenamefont {No{\'e}}\ \emph {et~al.}(2009)\citenamefont
  {No{\'e}}, \citenamefont {Sch{\"u}tte}, \citenamefont {Vanden-Eijnden},
  \citenamefont {Reich},\ and\ \citenamefont {Weikl}}]{noe2009constructing}%
  \BibitemOpen
  \bibfield  {author} {\bibinfo {author} {\bibfnamefont {F.}~\bibnamefont
  {No{\'e}}}, \bibinfo {author} {\bibfnamefont {C.}~\bibnamefont
  {Sch{\"u}tte}}, \bibinfo {author} {\bibfnamefont {E.}~\bibnamefont
  {Vanden-Eijnden}}, \bibinfo {author} {\bibfnamefont {L.}~\bibnamefont
  {Reich}}, \ and\ \bibinfo {author} {\bibfnamefont {T.~R.}\ \bibnamefont
  {Weikl}},\ }\bibfield  {title} {\enquote {\bibinfo {title} {Constructing the
  equilibrium ensemble of folding pathways from short off-equilibrium
  simulations},}\ }\href@noop {} {\bibfield  {journal} {\bibinfo  {journal}
  {Proceedings of the National Academy of Sciences}\ }\textbf {\bibinfo
  {volume} {106}},\ \bibinfo {pages} {19011--19016} (\bibinfo {year}
  {2009})}\BibitemShut {NoStop}%
\bibitem [{\citenamefont {Bowman}, \citenamefont {Pande},\ and\ \citenamefont
  {No{\'e}}(2013)}]{bowman2013introduction}%
  \BibitemOpen
  \bibfield  {author} {\bibinfo {author} {\bibfnamefont {G.~R.}\ \bibnamefont
  {Bowman}}, \bibinfo {author} {\bibfnamefont {V.~S.}\ \bibnamefont {Pande}}, \
  and\ \bibinfo {author} {\bibfnamefont {F.}~\bibnamefont {No{\'e}}},\
  }\href@noop {} {\emph {\bibinfo {title} {An Introduction to Markov State
  Models and their Application to Long Timescale Molecular Simulation}}},\
  Vol.\ \bibinfo {volume} {797}\ (\bibinfo  {publisher} {Springer Science \&
  Business Media},\ \bibinfo {year} {2013})\BibitemShut {NoStop}%
\bibitem [{\citenamefont {Wu}\ \emph {et~al.}(2017)\citenamefont {Wu},
  \citenamefont {N\"uske}, \citenamefont {Paul}, \citenamefont {Klus},
  \citenamefont {Koltai},\ and\ \citenamefont {No\'e}}]{wu_variational_2017}%
  \BibitemOpen
  \bibfield  {author} {\bibinfo {author} {\bibfnamefont {H.}~\bibnamefont
  {Wu}}, \bibinfo {author} {\bibfnamefont {F.}~\bibnamefont {N\"uske}},
  \bibinfo {author} {\bibfnamefont {F.}~\bibnamefont {Paul}}, \bibinfo {author}
  {\bibfnamefont {S.}~\bibnamefont {Klus}}, \bibinfo {author} {\bibfnamefont
  {P.}~\bibnamefont {Koltai}}, \ and\ \bibinfo {author} {\bibfnamefont
  {F.}~\bibnamefont {No\'e}},\ }\bibfield  {title} {\enquote {\bibinfo {title}
  {Variational {Koopman} models: {Slow} collective variables and molecular
  kinetics from short off-equilibrium simulations},}\ }\href {\doibase
  10.1063/1.4979344} {\bibfield  {journal} {\bibinfo  {journal} {Journal of
  Chemical Physics}\ }\textbf {\bibinfo {volume} {146}},\ \bibinfo {pages}
  {154104} (\bibinfo {year} {2017})}\BibitemShut {NoStop}%
\bibitem [{\citenamefont {Buchete}\ and\ \citenamefont
  {Hummer}(2008)}]{buchete2008coarse}%
  \BibitemOpen
  \bibfield  {author} {\bibinfo {author} {\bibfnamefont {N.-V.}\ \bibnamefont
  {Buchete}}\ and\ \bibinfo {author} {\bibfnamefont {G.}~\bibnamefont
  {Hummer}},\ }\bibfield  {title} {\enquote {\bibinfo {title} {Coarse master
  equations for peptide folding dynamics},}\ }\href@noop {} {\bibfield
  {journal} {\bibinfo  {journal} {Journal of Physical Chemistry B}\ }\textbf
  {\bibinfo {volume} {112}},\ \bibinfo {pages} {6057--6069} (\bibinfo {year}
  {2008})}\BibitemShut {NoStop}%
\bibitem [{\citenamefont {Thiede}\ \emph {et~al.}(2019)\citenamefont {Thiede},
  \citenamefont {Giannakis}, \citenamefont {Dinner},\ and\ \citenamefont
  {Weare}}]{thiede_galerkin_2019}%
  \BibitemOpen
  \bibfield  {author} {\bibinfo {author} {\bibfnamefont {E.~H.}\ \bibnamefont
  {Thiede}}, \bibinfo {author} {\bibfnamefont {D.}~\bibnamefont {Giannakis}},
  \bibinfo {author} {\bibfnamefont {A.~R.}\ \bibnamefont {Dinner}}, \ and\
  \bibinfo {author} {\bibfnamefont {J.}~\bibnamefont {Weare}},\ }\bibfield
  {title} {\enquote {\bibinfo {title} {Galerkin approximation of dynamical
  quantities using trajectory data},}\ }\href {\doibase 10.1063/1.5063730}
  {\bibfield  {journal} {\bibinfo  {journal} {Journal of Chemical Physics}\
  }\textbf {\bibinfo {volume} {150}},\ \bibinfo {pages} {244111} (\bibinfo
  {year} {2019})}\BibitemShut {NoStop}%
\bibitem [{\citenamefont {Strahan}\ \emph {et~al.}(2021)\citenamefont
  {Strahan}, \citenamefont {Antoszewski}, \citenamefont {Lorpaiboon},
  \citenamefont {Vani}, \citenamefont {Weare},\ and\ \citenamefont
  {Dinner}}]{strahan2021long}%
  \BibitemOpen
  \bibfield  {author} {\bibinfo {author} {\bibfnamefont {J.}~\bibnamefont
  {Strahan}}, \bibinfo {author} {\bibfnamefont {A.}~\bibnamefont
  {Antoszewski}}, \bibinfo {author} {\bibfnamefont {C.}~\bibnamefont
  {Lorpaiboon}}, \bibinfo {author} {\bibfnamefont {B.~P.}\ \bibnamefont
  {Vani}}, \bibinfo {author} {\bibfnamefont {J.}~\bibnamefont {Weare}}, \ and\
  \bibinfo {author} {\bibfnamefont {A.~R.}\ \bibnamefont {Dinner}},\ }\bibfield
   {title} {\enquote {\bibinfo {title} {Long-time-scale predictions from
  short-trajectory data: A benchmark analysis of the trp-cage miniprotein},}\
  }\href@noop {} {\bibfield  {journal} {\bibinfo  {journal} {Journal of
  chemical theory and computation}\ }\textbf {\bibinfo {volume} {17}},\
  \bibinfo {pages} {2948--2963} (\bibinfo {year} {2021})}\BibitemShut {NoStop}%
\bibitem [{\citenamefont {Russo}\ \emph {et~al.}(2021)\citenamefont {Russo},
  \citenamefont {Copperman}, \citenamefont {Aristoff}, \citenamefont
  {Simpson},\ and\ \citenamefont {Zuckerman}}]{russo2021unbiased}%
  \BibitemOpen
  \bibfield  {author} {\bibinfo {author} {\bibfnamefont {J.~D.}\ \bibnamefont
  {Russo}}, \bibinfo {author} {\bibfnamefont {J.}~\bibnamefont {Copperman}},
  \bibinfo {author} {\bibfnamefont {D.}~\bibnamefont {Aristoff}}, \bibinfo
  {author} {\bibfnamefont {G.}~\bibnamefont {Simpson}}, \ and\ \bibinfo
  {author} {\bibfnamefont {D.~M.}\ \bibnamefont {Zuckerman}},\ }\bibfield
  {title} {\enquote {\bibinfo {title} {Unbiased estimation of equilibrium,
  rates, and committors from {Markov} state model analysis},}\ }\href@noop {}
  {\bibfield  {journal} {\bibinfo  {journal} {arXiv:2105.13402}\ } (\bibinfo
  {year} {2021})}\BibitemShut {NoStop}%
\bibitem [{\citenamefont {Elber}\ \emph {et~al.}(2017)\citenamefont {Elber},
  \citenamefont {Bello-Rivas}, \citenamefont {Ma}, \citenamefont {Cardenas},\
  and\ \citenamefont {Fathizadeh}}]{elber2017calculating}%
  \BibitemOpen
  \bibfield  {author} {\bibinfo {author} {\bibfnamefont {R.}~\bibnamefont
  {Elber}}, \bibinfo {author} {\bibfnamefont {J.~M.}\ \bibnamefont
  {Bello-Rivas}}, \bibinfo {author} {\bibfnamefont {P.}~\bibnamefont {Ma}},
  \bibinfo {author} {\bibfnamefont {A.~E.}\ \bibnamefont {Cardenas}}, \ and\
  \bibinfo {author} {\bibfnamefont {A.}~\bibnamefont {Fathizadeh}},\ }\bibfield
   {title} {\enquote {\bibinfo {title} {Calculating iso-committor surfaces as
  optimal reaction coordinates with milestoning},}\ }\href@noop {} {\bibfield
  {journal} {\bibinfo  {journal} {Entropy}\ }\textbf {\bibinfo {volume} {19}},\
  \bibinfo {pages} {219} (\bibinfo {year} {2017})}\BibitemShut {NoStop}%
\bibitem [{\citenamefont {Faradjian}\ and\ \citenamefont
  {Elber}(2004)}]{faradjian2004computing}%
  \BibitemOpen
  \bibfield  {author} {\bibinfo {author} {\bibfnamefont {A.~K.}\ \bibnamefont
  {Faradjian}}\ and\ \bibinfo {author} {\bibfnamefont {R.}~\bibnamefont
  {Elber}},\ }\bibfield  {title} {\enquote {\bibinfo {title} {Computing time
  scales from reaction coordinates by milestoning},}\ }\href@noop {} {\bibfield
   {journal} {\bibinfo  {journal} {Journal of Chemical Physics}\ }\textbf
  {\bibinfo {volume} {120}},\ \bibinfo {pages} {10880--10889} (\bibinfo {year}
  {2004})}\BibitemShut {NoStop}%
\bibitem [{\citenamefont {Elber}(2020)}]{elber2020milestoning}%
  \BibitemOpen
  \bibfield  {author} {\bibinfo {author} {\bibfnamefont {R.}~\bibnamefont
  {Elber}},\ }\bibfield  {title} {\enquote {\bibinfo {title} {Milestoning: An
  efficient approach for atomically detailed simulations of kinetics in
  biophysics},}\ }\href@noop {} {\bibfield  {journal} {\bibinfo  {journal}
  {Annual Review of Biophysics}\ }\textbf {\bibinfo {volume} {49}},\ \bibinfo
  {pages} {69--85} (\bibinfo {year} {2020})}\BibitemShut {NoStop}%
\bibitem [{\citenamefont {Dellago}\ \emph {et~al.}(1998)\citenamefont
  {Dellago}, \citenamefont {Bolhuis}, \citenamefont {Csajka},\ and\
  \citenamefont {Chandler}}]{dellago1998transition}%
  \BibitemOpen
  \bibfield  {author} {\bibinfo {author} {\bibfnamefont {C.}~\bibnamefont
  {Dellago}}, \bibinfo {author} {\bibfnamefont {P.~G.}\ \bibnamefont
  {Bolhuis}}, \bibinfo {author} {\bibfnamefont {F.~S.}\ \bibnamefont {Csajka}},
  \ and\ \bibinfo {author} {\bibfnamefont {D.}~\bibnamefont {Chandler}},\
  }\bibfield  {title} {\enquote {\bibinfo {title} {Transition path sampling and
  the calculation of rate constants},}\ }\href@noop {} {\bibfield  {journal}
  {\bibinfo  {journal} {Journal of Chemical Physics}\ }\textbf {\bibinfo
  {volume} {108}},\ \bibinfo {pages} {1964--1977} (\bibinfo {year}
  {1998})}\BibitemShut {NoStop}%
\bibitem [{\citenamefont {Dellago}, \citenamefont {Bolhuis},\ and\
  \citenamefont {Chandler}(1998)}]{dellago1998efficient}%
  \BibitemOpen
  \bibfield  {author} {\bibinfo {author} {\bibfnamefont {C.}~\bibnamefont
  {Dellago}}, \bibinfo {author} {\bibfnamefont {P.~G.}\ \bibnamefont
  {Bolhuis}}, \ and\ \bibinfo {author} {\bibfnamefont {D.}~\bibnamefont
  {Chandler}},\ }\bibfield  {title} {\enquote {\bibinfo {title} {Efficient
  transition path sampling: Application to {L}ennard-{J}ones cluster
  rearrangements},}\ }\href@noop {} {\bibfield  {journal} {\bibinfo  {journal}
  {Journal of Chemical Physics}\ }\textbf {\bibinfo {volume} {108}},\ \bibinfo
  {pages} {9236--9245} (\bibinfo {year} {1998})}\BibitemShut {NoStop}%
\bibitem [{\citenamefont {Dellago}\ \emph {et~al.}(2002)\citenamefont
  {Dellago}, \citenamefont {Bolhuis}, \citenamefont {Geissler} \emph
  {et~al.}}]{dellago2002transition}%
  \BibitemOpen
  \bibfield  {author} {\bibinfo {author} {\bibfnamefont {C.}~\bibnamefont
  {Dellago}}, \bibinfo {author} {\bibfnamefont {P.}~\bibnamefont {Bolhuis}},
  \bibinfo {author} {\bibfnamefont {P.~L.}\ \bibnamefont {Geissler}},  \emph
  {et~al.},\ }\bibfield  {title} {\enquote {\bibinfo {title} {Transition path
  sampling},}\ }\href@noop {} {\bibfield  {journal} {\bibinfo  {journal}
  {Advances in Chemical Physics}\ }\textbf {\bibinfo {volume} {123}} (\bibinfo
  {year} {2002})}\BibitemShut {NoStop}%
\bibitem [{\citenamefont {Huber}\ and\ \citenamefont
  {Kim}(1996)}]{huber_weighted-ensemble_1996}%
  \BibitemOpen
  \bibfield  {author} {\bibinfo {author} {\bibfnamefont {G.~A.}\ \bibnamefont
  {Huber}}\ and\ \bibinfo {author} {\bibfnamefont {S.}~\bibnamefont {Kim}},\
  }\bibfield  {title} {\enquote {\bibinfo {title} {Weighted-ensemble {Brownian}
  dynamics simulations for protein association reactions.}}\ }\href
  {https://www.ncbi.nlm.nih.gov/pmc/articles/PMC1224912/} {\bibfield  {journal}
  {\bibinfo  {journal} {Biophysical Journal}\ }\textbf {\bibinfo {volume}
  {70}},\ \bibinfo {pages} {97--110} (\bibinfo {year} {1996})}\BibitemShut
  {NoStop}%
\bibitem [{\citenamefont {Zhang}, \citenamefont {Jasnow},\ and\ \citenamefont
  {Zuckerman}(2007)}]{zhang2007efficient}%
  \BibitemOpen
  \bibfield  {author} {\bibinfo {author} {\bibfnamefont {B.~W.}\ \bibnamefont
  {Zhang}}, \bibinfo {author} {\bibfnamefont {D.}~\bibnamefont {Jasnow}}, \
  and\ \bibinfo {author} {\bibfnamefont {D.~M.}\ \bibnamefont {Zuckerman}},\
  }\bibfield  {title} {\enquote {\bibinfo {title} {Efficient and verified
  simulation of a path ensemble for conformational change in a united-residue
  model of calmodulin},}\ }\href@noop {} {\bibfield  {journal} {\bibinfo
  {journal} {Proceedings of the National Academy of Sciences}\ }\textbf
  {\bibinfo {volume} {104}},\ \bibinfo {pages} {18043--18048} (\bibinfo {year}
  {2007})}\BibitemShut {NoStop}%
\bibitem [{\citenamefont {Zuckerman}\ and\ \citenamefont
  {Chong}(2017)}]{zuckerman2017weighted}%
  \BibitemOpen
  \bibfield  {author} {\bibinfo {author} {\bibfnamefont {D.~M.}\ \bibnamefont
  {Zuckerman}}\ and\ \bibinfo {author} {\bibfnamefont {L.~T.}\ \bibnamefont
  {Chong}},\ }\bibfield  {title} {\enquote {\bibinfo {title} {Weighted ensemble
  simulation: review of methodology, applications, and software},}\ }\href@noop
  {} {\bibfield  {journal} {\bibinfo  {journal} {Annual Review of Biophysics}\
  }\textbf {\bibinfo {volume} {46}},\ \bibinfo {pages} {43--57} (\bibinfo
  {year} {2017})}\BibitemShut {NoStop}%
\bibitem [{\citenamefont {Bolhuis}(2003)}]{bolhuis2003transition}%
  \BibitemOpen
  \bibfield  {author} {\bibinfo {author} {\bibfnamefont {P.~G.}\ \bibnamefont
  {Bolhuis}},\ }\bibfield  {title} {\enquote {\bibinfo {title} {Transition-path
  sampling of $\beta$-hairpin folding},}\ }\href@noop {} {\bibfield  {journal}
  {\bibinfo  {journal} {Proceedings of the National Academy of Sciences}\
  }\textbf {\bibinfo {volume} {100}},\ \bibinfo {pages} {12129--12134}
  (\bibinfo {year} {2003})}\BibitemShut {NoStop}%
\bibitem [{\citenamefont {van Erp}, \citenamefont {Moroni},\ and\ \citenamefont
  {Bolhuis}(2003)}]{van2003novel}%
  \BibitemOpen
  \bibfield  {author} {\bibinfo {author} {\bibfnamefont {T.~S.}\ \bibnamefont
  {van Erp}}, \bibinfo {author} {\bibfnamefont {D.}~\bibnamefont {Moroni}}, \
  and\ \bibinfo {author} {\bibfnamefont {P.~G.}\ \bibnamefont {Bolhuis}},\
  }\bibfield  {title} {\enquote {\bibinfo {title} {A novel path sampling method
  for the calculation of rate constants},}\ }\href@noop {} {\bibfield
  {journal} {\bibinfo  {journal} {Journal of Chemical Physics}\ }\textbf
  {\bibinfo {volume} {118}},\ \bibinfo {pages} {7762--7774} (\bibinfo {year}
  {2003})}\BibitemShut {NoStop}%
\bibitem [{\citenamefont {Allen}, \citenamefont {Warren},\ and\ \citenamefont
  {Ten~Wolde}(2005)}]{allen2005sampling}%
  \BibitemOpen
  \bibfield  {author} {\bibinfo {author} {\bibfnamefont {R.~J.}\ \bibnamefont
  {Allen}}, \bibinfo {author} {\bibfnamefont {P.~B.}\ \bibnamefont {Warren}}, \
  and\ \bibinfo {author} {\bibfnamefont {P.~R.}\ \bibnamefont {Ten~Wolde}},\
  }\bibfield  {title} {\enquote {\bibinfo {title} {Sampling rare switching
  events in biochemical networks},}\ }\href@noop {} {\bibfield  {journal}
  {\bibinfo  {journal} {Physical Review Letters}\ }\textbf {\bibinfo {volume}
  {94}},\ \bibinfo {pages} {018104} (\bibinfo {year} {2005})}\BibitemShut
  {NoStop}%
\bibitem [{\citenamefont {Allen}, \citenamefont {Valeriani},\ and\
  \citenamefont {ten Wolde}(2009)}]{allen2009forward}%
  \BibitemOpen
  \bibfield  {author} {\bibinfo {author} {\bibfnamefont {R.~J.}\ \bibnamefont
  {Allen}}, \bibinfo {author} {\bibfnamefont {C.}~\bibnamefont {Valeriani}}, \
  and\ \bibinfo {author} {\bibfnamefont {P.~R.}\ \bibnamefont {ten Wolde}},\
  }\bibfield  {title} {\enquote {\bibinfo {title} {Forward flux sampling for
  rare event simulations},}\ }\href@noop {} {\bibfield  {journal} {\bibinfo
  {journal} {Journal of Physics: Condensed Matter}\ }\textbf {\bibinfo {volume}
  {21}},\ \bibinfo {pages} {463102} (\bibinfo {year} {2009})}\BibitemShut
  {NoStop}%
\bibitem [{\citenamefont {Warmflash}, \citenamefont {Bhimalapuram},\ and\
  \citenamefont {Dinner}(2007)}]{warmflash2007umbrella}%
  \BibitemOpen
  \bibfield  {author} {\bibinfo {author} {\bibfnamefont {A.}~\bibnamefont
  {Warmflash}}, \bibinfo {author} {\bibfnamefont {P.}~\bibnamefont
  {Bhimalapuram}}, \ and\ \bibinfo {author} {\bibfnamefont {A.~R.}\
  \bibnamefont {Dinner}},\ }\bibfield  {title} {\enquote {\bibinfo {title}
  {Umbrella sampling for nonequilibrium processes},}\ }\href@noop {} {\bibfield
   {journal} {\bibinfo  {journal} {Journal of Chemical Physics}\ }\textbf
  {\bibinfo {volume} {127}},\ \bibinfo {pages} {114109} (\bibinfo {year}
  {2007})}\BibitemShut {NoStop}%
\bibitem [{\citenamefont {Dickson}, \citenamefont {Warmflash},\ and\
  \citenamefont {Dinner}(2009{\natexlab{a}})}]{dickson2009separating}%
  \BibitemOpen
  \bibfield  {author} {\bibinfo {author} {\bibfnamefont {A.}~\bibnamefont
  {Dickson}}, \bibinfo {author} {\bibfnamefont {A.}~\bibnamefont {Warmflash}},
  \ and\ \bibinfo {author} {\bibfnamefont {A.~R.}\ \bibnamefont {Dinner}},\
  }\bibfield  {title} {\enquote {\bibinfo {title} {Separating forward and
  backward pathways in nonequilibrium umbrella sampling},}\ }\href@noop {}
  {\bibfield  {journal} {\bibinfo  {journal} {Journal of Chemical Physics}\
  }\textbf {\bibinfo {volume} {131}},\ \bibinfo {pages} {154104} (\bibinfo
  {year} {2009}{\natexlab{a}})}\BibitemShut {NoStop}%
\bibitem [{\citenamefont {Vanden-Eijnden}\ and\ \citenamefont
  {Venturoli}(2009)}]{vanden2009exact}%
  \BibitemOpen
  \bibfield  {author} {\bibinfo {author} {\bibfnamefont {E.}~\bibnamefont
  {Vanden-Eijnden}}\ and\ \bibinfo {author} {\bibfnamefont {M.}~\bibnamefont
  {Venturoli}},\ }\bibfield  {title} {\enquote {\bibinfo {title} {Exact rate
  calculations by trajectory parallelization and tilting},}\ }\href@noop {}
  {\bibfield  {journal} {\bibinfo  {journal} {Journal of Chemical Physics}\
  }\textbf {\bibinfo {volume} {131}},\ \bibinfo {pages} {044120} (\bibinfo
  {year} {2009})}\BibitemShut {NoStop}%
\bibitem [{\citenamefont {Dickson}\ and\ \citenamefont
  {Dinner}(2010)}]{dickson2010enhanced}%
  \BibitemOpen
  \bibfield  {author} {\bibinfo {author} {\bibfnamefont {A.}~\bibnamefont
  {Dickson}}\ and\ \bibinfo {author} {\bibfnamefont {A.~R.}\ \bibnamefont
  {Dinner}},\ }\bibfield  {title} {\enquote {\bibinfo {title} {Enhanced
  sampling of nonequilibrium steady states},}\ }\href@noop {} {\bibfield
  {journal} {\bibinfo  {journal} {Annual Review of Physical Chemistry}\
  }\textbf {\bibinfo {volume} {61}},\ \bibinfo {pages} {441--459} (\bibinfo
  {year} {2010})}\BibitemShut {NoStop}%
\bibitem [{\citenamefont {Dinner}\ \emph {et~al.}(2020)\citenamefont {Dinner},
  \citenamefont {Thiede}, \citenamefont {{Van Koten}},\ and\ \citenamefont
  {Weare}}]{dinner2020stratification}%
  \BibitemOpen
  \bibfield  {author} {\bibinfo {author} {\bibfnamefont {A.~R.}\ \bibnamefont
  {Dinner}}, \bibinfo {author} {\bibfnamefont {E.~H.}\ \bibnamefont {Thiede}},
  \bibinfo {author} {\bibfnamefont {B.}~\bibnamefont {{Van Koten}}}, \ and\
  \bibinfo {author} {\bibfnamefont {J.}~\bibnamefont {Weare}},\ }\bibfield
  {title} {\enquote {\bibinfo {title} {Stratification as a general variance
  reduction method for {Markov chain Monte Carlo}},}\ }\href@noop {} {\bibfield
   {journal} {\bibinfo  {journal} {SIAM/ASA Journal on Uncertainty
  Quantification}\ }\textbf {\bibinfo {volume} {8}},\ \bibinfo {pages}
  {1139--1188} (\bibinfo {year} {2020})}\BibitemShut {NoStop}%
\bibitem [{\citenamefont {Bello-Rivas}\ and\ \citenamefont
  {Elber}(2015)}]{bello2015exact}%
  \BibitemOpen
  \bibfield  {author} {\bibinfo {author} {\bibfnamefont {J.~M.}\ \bibnamefont
  {Bello-Rivas}}\ and\ \bibinfo {author} {\bibfnamefont {R.}~\bibnamefont
  {Elber}},\ }\bibfield  {title} {\enquote {\bibinfo {title} {Exact
  milestoning},}\ }\href@noop {} {\bibfield  {journal} {\bibinfo  {journal}
  {Journal of Chemical Physics}\ }\textbf {\bibinfo {volume} {142}},\ \bibinfo
  {pages} {03B602\_1} (\bibinfo {year} {2015})}\BibitemShut {NoStop}%
\bibitem [{\citenamefont {Bhatt}, \citenamefont {Zhang},\ and\ \citenamefont
  {Zuckerman}(2010)}]{bhatt2010steady}%
  \BibitemOpen
  \bibfield  {author} {\bibinfo {author} {\bibfnamefont {D.}~\bibnamefont
  {Bhatt}}, \bibinfo {author} {\bibfnamefont {B.~W.}\ \bibnamefont {Zhang}}, \
  and\ \bibinfo {author} {\bibfnamefont {D.~M.}\ \bibnamefont {Zuckerman}},\
  }\bibfield  {title} {\enquote {\bibinfo {title} {Steady-state simulations
  using weighted ensemble path sampling},}\ }\href@noop {} {\bibfield
  {journal} {\bibinfo  {journal} {Journal of Chemical Physics}\ }\textbf
  {\bibinfo {volume} {133}},\ \bibinfo {pages} {014110} (\bibinfo {year}
  {2010})}\BibitemShut {NoStop}%
\bibitem [{\citenamefont {Dinner}\ \emph {et~al.}(2018)\citenamefont {Dinner},
  \citenamefont {Mattingly}, \citenamefont {Tempkin}, \citenamefont {{Van
  Koten}},\ and\ \citenamefont {Weare}}]{dinner2018trajectory}%
  \BibitemOpen
  \bibfield  {author} {\bibinfo {author} {\bibfnamefont {A.~R.}\ \bibnamefont
  {Dinner}}, \bibinfo {author} {\bibfnamefont {J.~C.}\ \bibnamefont
  {Mattingly}}, \bibinfo {author} {\bibfnamefont {J.~O.}\ \bibnamefont
  {Tempkin}}, \bibinfo {author} {\bibfnamefont {B.}~\bibnamefont {{Van
  Koten}}}, \ and\ \bibinfo {author} {\bibfnamefont {J.}~\bibnamefont
  {Weare}},\ }\bibfield  {title} {\enquote {\bibinfo {title} {Trajectory
  stratification of stochastic dynamics},}\ }\href@noop {} {\bibfield
  {journal} {\bibinfo  {journal} {SIAM Review}\ }\textbf {\bibinfo {volume}
  {60}},\ \bibinfo {pages} {909--938} (\bibinfo {year} {2018})}\BibitemShut
  {NoStop}%
\bibitem [{\citenamefont {Lorpaiboon}, \citenamefont {Weare},\ and\
  \citenamefont {Dinner}(2022)}]{lorpaiboonATPT}%
  \BibitemOpen
  \bibfield  {author} {\bibinfo {author} {\bibfnamefont {C.}~\bibnamefont
  {Lorpaiboon}}, \bibinfo {author} {\bibfnamefont {J.}~\bibnamefont {Weare}}, \
  and\ \bibinfo {author} {\bibfnamefont {A.~R.}\ \bibnamefont {Dinner}},\
  }\bibfield  {title} {\enquote {\bibinfo {title} {Augmented transition path
  theory for sequences of events},}\ }\href@noop {} {\bibfield  {journal}
  {\bibinfo  {journal} {in preparation}\ } (\bibinfo {year}
  {2022})}\BibitemShut {NoStop}%
\bibitem [{\citenamefont {Metzner}, \citenamefont {Schütte},\ and\
  \citenamefont {Vanden-Eijnden}(2009)}]{metzner_transition_2009}%
  \BibitemOpen
  \bibfield  {author} {\bibinfo {author} {\bibfnamefont {P.}~\bibnamefont
  {Metzner}}, \bibinfo {author} {\bibfnamefont {C.}~\bibnamefont {Schütte}}, \
  and\ \bibinfo {author} {\bibfnamefont {E.}~\bibnamefont {Vanden-Eijnden}},\
  }\bibfield  {title} {\enquote {\bibinfo {title} {Transition path theory for
  {Markov} jump processes},}\ }\href@noop {} {\bibfield  {journal} {\bibinfo
  {journal} {Multiscale Modeling \& Simulation}\ }\textbf {\bibinfo {volume}
  {7}},\ \bibinfo {pages} {1192--1219} (\bibinfo {year} {2009})}\BibitemShut
  {NoStop}%
\bibitem [{\citenamefont {Torrie}\ and\ \citenamefont
  {Valleau}(1977)}]{torrie1977nonphysical}%
  \BibitemOpen
  \bibfield  {author} {\bibinfo {author} {\bibfnamefont {G.~M.}\ \bibnamefont
  {Torrie}}\ and\ \bibinfo {author} {\bibfnamefont {J.~P.}\ \bibnamefont
  {Valleau}},\ }\bibfield  {title} {\enquote {\bibinfo {title} {Nonphysical
  sampling distributions in {M}onte {C}arlo free-energy estimation: Umbrella
  sampling},}\ }\href@noop {} {\bibfield  {journal} {\bibinfo  {journal}
  {Journal of Computational Physics}\ }\textbf {\bibinfo {volume} {23}},\
  \bibinfo {pages} {187--199} (\bibinfo {year} {1977})}\BibitemShut {NoStop}%
\bibitem [{\citenamefont {Pangali}, \citenamefont {Rao},\ and\ \citenamefont
  {Berne}(1979)}]{pangali1979monte}%
  \BibitemOpen
  \bibfield  {author} {\bibinfo {author} {\bibfnamefont {C.}~\bibnamefont
  {Pangali}}, \bibinfo {author} {\bibfnamefont {M.}~\bibnamefont {Rao}}, \ and\
  \bibinfo {author} {\bibfnamefont {B.}~\bibnamefont {Berne}},\ }\bibfield
  {title} {\enquote {\bibinfo {title} {A {M}onte {C}arlo simulation of the
  hydrophobic interaction},}\ }\href@noop {} {\bibfield  {journal} {\bibinfo
  {journal} {Journal of Chemical Physics}\ }\textbf {\bibinfo {volume} {71}},\
  \bibinfo {pages} {2975--2981} (\bibinfo {year} {1979})}\BibitemShut {NoStop}%
\bibitem [{\citenamefont {Frenkel}\ and\ \citenamefont
  {Smit}(2001)}]{frenkel2001understanding}%
  \BibitemOpen
  \bibfield  {author} {\bibinfo {author} {\bibfnamefont {D.}~\bibnamefont
  {Frenkel}}\ and\ \bibinfo {author} {\bibfnamefont {B.}~\bibnamefont {Smit}},\
  }\href@noop {} {\emph {\bibinfo {title} {Understanding Molecular Simulation:
  from Algorithms to Applications}}}\ (\bibinfo  {publisher} {Elsevier},\
  \bibinfo {year} {2001})\BibitemShut {NoStop}%
\bibitem [{\citenamefont {Thiede}\ \emph {et~al.}(2016)\citenamefont {Thiede},
  \citenamefont {{Van Koten}}, \citenamefont {Weare},\ and\ \citenamefont
  {Dinner}}]{thiede2016eigenvector}%
  \BibitemOpen
  \bibfield  {author} {\bibinfo {author} {\bibfnamefont {E.~H.}\ \bibnamefont
  {Thiede}}, \bibinfo {author} {\bibfnamefont {B.}~\bibnamefont {{Van Koten}}},
  \bibinfo {author} {\bibfnamefont {J.}~\bibnamefont {Weare}}, \ and\ \bibinfo
  {author} {\bibfnamefont {A.~R.}\ \bibnamefont {Dinner}},\ }\bibfield  {title}
  {\enquote {\bibinfo {title} {Eigenvector method for umbrella sampling enables
  error analysis},}\ }\href@noop {} {\bibfield  {journal} {\bibinfo  {journal}
  {Journal of Chemical Physics}\ }\textbf {\bibinfo {volume} {145}},\ \bibinfo
  {pages} {084115} (\bibinfo {year} {2016})}\BibitemShut {NoStop}%
\bibitem [{\citenamefont {Dickson}, \citenamefont {Warmflash},\ and\
  \citenamefont {Dinner}(2009{\natexlab{b}})}]{dickson2009nonequilibrium}%
  \BibitemOpen
  \bibfield  {author} {\bibinfo {author} {\bibfnamefont {A.}~\bibnamefont
  {Dickson}}, \bibinfo {author} {\bibfnamefont {A.}~\bibnamefont {Warmflash}},
  \ and\ \bibinfo {author} {\bibfnamefont {A.~R.}\ \bibnamefont {Dinner}},\
  }\bibfield  {title} {\enquote {\bibinfo {title} {Nonequilibrium umbrella
  sampling in spaces of many order parameters},}\ }\href@noop {} {\bibfield
  {journal} {\bibinfo  {journal} {Journal of Chemical Physics}\ }\textbf
  {\bibinfo {volume} {130}},\ \bibinfo {pages} {02B605} (\bibinfo {year}
  {2009}{\natexlab{b}})}\BibitemShut {NoStop}%
\bibitem [{\citenamefont {Guttenberg}, \citenamefont {Dinner},\ and\
  \citenamefont {Weare}(2012)}]{guttenberg2012steered}%
  \BibitemOpen
  \bibfield  {author} {\bibinfo {author} {\bibfnamefont {N.}~\bibnamefont
  {Guttenberg}}, \bibinfo {author} {\bibfnamefont {A.~R.}\ \bibnamefont
  {Dinner}}, \ and\ \bibinfo {author} {\bibfnamefont {J.}~\bibnamefont
  {Weare}},\ }\bibfield  {title} {\enquote {\bibinfo {title} {Steered
  transition path sampling},}\ }\href@noop {} {\bibfield  {journal} {\bibinfo
  {journal} {Journal of Chemical Physics}\ }\textbf {\bibinfo {volume} {136}},\
  \bibinfo {pages} {234103} (\bibinfo {year} {2012})}\BibitemShut {NoStop}%
\bibitem [{\citenamefont {MacKerell}\ \emph {et~al.}(1998)\citenamefont
  {MacKerell}, \citenamefont {Bashford}, \citenamefont {Bellott}, \citenamefont
  {Dunbrack}, \citenamefont {Evanseck}, \citenamefont {Field}, \citenamefont
  {Fischer}, \citenamefont {Gao}, \citenamefont {Guo}, \citenamefont {Ha},
  \citenamefont {Joseph-McCarthy}, \citenamefont {Kuchnir}, \citenamefont
  {Kuczera}, \citenamefont {Lau}, \citenamefont {Mattos}, \citenamefont
  {Michnick}, \citenamefont {Ngo}, \citenamefont {Nguyen}, \citenamefont
  {Prodhom}, \citenamefont {Reiher}, \citenamefont {Roux}, \citenamefont
  {Schlenkrich}, \citenamefont {Smith}, \citenamefont {Stote}, \citenamefont
  {Straub}, \citenamefont {Watanabe}, \citenamefont
  {Wi{\'{o}}rkiewicz-Kuczera}, \citenamefont {Yin},\ and\ \citenamefont
  {Karplus}}]{MacKerell1998}%
  \BibitemOpen
  \bibfield  {author} {\bibinfo {author} {\bibfnamefont {A.~D.}\ \bibnamefont
  {MacKerell}}, \bibinfo {author} {\bibfnamefont {D.}~\bibnamefont {Bashford}},
  \bibinfo {author} {\bibfnamefont {M.}~\bibnamefont {Bellott}}, \bibinfo
  {author} {\bibfnamefont {R.~L.}\ \bibnamefont {Dunbrack}}, \bibinfo {author}
  {\bibfnamefont {J.~D.}\ \bibnamefont {Evanseck}}, \bibinfo {author}
  {\bibfnamefont {M.~J.}\ \bibnamefont {Field}}, \bibinfo {author}
  {\bibfnamefont {S.}~\bibnamefont {Fischer}}, \bibinfo {author} {\bibfnamefont
  {J.}~\bibnamefont {Gao}}, \bibinfo {author} {\bibfnamefont {H.}~\bibnamefont
  {Guo}}, \bibinfo {author} {\bibfnamefont {S.}~\bibnamefont {Ha}}, \bibinfo
  {author} {\bibfnamefont {D.}~\bibnamefont {Joseph-McCarthy}}, \bibinfo
  {author} {\bibfnamefont {L.}~\bibnamefont {Kuchnir}}, \bibinfo {author}
  {\bibfnamefont {K.}~\bibnamefont {Kuczera}}, \bibinfo {author} {\bibfnamefont
  {F.~T.~K.}\ \bibnamefont {Lau}}, \bibinfo {author} {\bibfnamefont
  {C.}~\bibnamefont {Mattos}}, \bibinfo {author} {\bibfnamefont
  {S.}~\bibnamefont {Michnick}}, \bibinfo {author} {\bibfnamefont
  {T.}~\bibnamefont {Ngo}}, \bibinfo {author} {\bibfnamefont {D.~T.}\
  \bibnamefont {Nguyen}}, \bibinfo {author} {\bibfnamefont {B.}~\bibnamefont
  {Prodhom}}, \bibinfo {author} {\bibfnamefont {W.~E.}\ \bibnamefont {Reiher}},
  \bibinfo {author} {\bibfnamefont {B.}~\bibnamefont {Roux}}, \bibinfo {author}
  {\bibfnamefont {M.}~\bibnamefont {Schlenkrich}}, \bibinfo {author}
  {\bibfnamefont {J.~C.}\ \bibnamefont {Smith}}, \bibinfo {author}
  {\bibfnamefont {R.}~\bibnamefont {Stote}}, \bibinfo {author} {\bibfnamefont
  {J.}~\bibnamefont {Straub}}, \bibinfo {author} {\bibfnamefont
  {M.}~\bibnamefont {Watanabe}}, \bibinfo {author} {\bibfnamefont
  {J.}~\bibnamefont {Wi{\'{o}}rkiewicz-Kuczera}}, \bibinfo {author}
  {\bibfnamefont {D.}~\bibnamefont {Yin}}, \ and\ \bibinfo {author}
  {\bibfnamefont {M.}~\bibnamefont {Karplus}},\ }\bibfield  {title} {\enquote
  {\bibinfo {title} {All-atom empirical potential for molecular modeling and
  dynamics studies of proteins},}\ }\href {\doibase 10.1021/jp973084f}
  {\bibfield  {journal} {\bibinfo  {journal} {Journal of Physical Chemistry B}\
  }\textbf {\bibinfo {volume} {102}},\ \bibinfo {pages} {3586--3616} (\bibinfo
  {year} {1998})}\BibitemShut {NoStop}%
\bibitem [{\citenamefont {Best}\ \emph {et~al.}(2012)\citenamefont {Best},
  \citenamefont {Zhu}, \citenamefont {Shim}, \citenamefont {Lopes},
  \citenamefont {Mittal}, \citenamefont {Feig},\ and\ \citenamefont
  {MacKerell}}]{Best2012}%
  \BibitemOpen
  \bibfield  {author} {\bibinfo {author} {\bibfnamefont {R.~B.}\ \bibnamefont
  {Best}}, \bibinfo {author} {\bibfnamefont {X.}~\bibnamefont {Zhu}}, \bibinfo
  {author} {\bibfnamefont {J.}~\bibnamefont {Shim}}, \bibinfo {author}
  {\bibfnamefont {P.~E.~M.}\ \bibnamefont {Lopes}}, \bibinfo {author}
  {\bibfnamefont {J.}~\bibnamefont {Mittal}}, \bibinfo {author} {\bibfnamefont
  {M.}~\bibnamefont {Feig}}, \ and\ \bibinfo {author} {\bibfnamefont {A.~D.}\
  \bibnamefont {MacKerell}},\ }\bibfield  {title} {\enquote {\bibinfo {title}
  {Optimization of the additive {CHARMM} all-atom protein force field targeting
  improved sampling of the backbone $\phi$, $\psi$ and side-chain $\chi_1$ and
  $\chi_2$ dihedral angles},}\ }\href {\doibase 10.1021/ct300400x} {\bibfield
  {journal} {\bibinfo  {journal} {Journal of Chemical Theory and Computation}\
  }\textbf {\bibinfo {volume} {8}},\ \bibinfo {pages} {3257--3273} (\bibinfo
  {year} {2012})}\BibitemShut {NoStop}%
\bibitem [{\citenamefont {Huang}\ \emph {et~al.}(2017)\citenamefont {Huang},
  \citenamefont {Rauscher}, \citenamefont {Nawrocki}, \citenamefont {Ran},
  \citenamefont {Feig}, \citenamefont {de~Groot}, \citenamefont
  {Grubm{\"{u}}ller},\ and\ \citenamefont {Jr}}]{Huang2017}%
  \BibitemOpen
  \bibfield  {author} {\bibinfo {author} {\bibfnamefont {J.}~\bibnamefont
  {Huang}}, \bibinfo {author} {\bibfnamefont {S.}~\bibnamefont {Rauscher}},
  \bibinfo {author} {\bibfnamefont {G.}~\bibnamefont {Nawrocki}}, \bibinfo
  {author} {\bibfnamefont {T.}~\bibnamefont {Ran}}, \bibinfo {author}
  {\bibfnamefont {M.}~\bibnamefont {Feig}}, \bibinfo {author} {\bibfnamefont
  {B.~L.}\ \bibnamefont {de~Groot}}, \bibinfo {author} {\bibfnamefont
  {H.}~\bibnamefont {Grubm{\"{u}}ller}}, \ and\ \bibinfo {author}
  {\bibfnamefont {A.~D.~M.}\ \bibnamefont {Jr}},\ }\bibfield  {title} {\enquote
  {\bibinfo {title} {{CHARMM36m}: An improved force field for folded and
  intrinsically disordered proteins},}\ }\href {\doibase
  10.1038/nmeth.4067.CHARMM36m} {\bibfield  {journal} {\bibinfo  {journal}
  {Nature Methods}\ }\textbf {\bibinfo {volume} {14}},\ \bibinfo {pages}
  {71--73} (\bibinfo {year} {2017})}\BibitemShut {NoStop}%
\bibitem [{\citenamefont {Goga}\ \emph {et~al.}(2012)\citenamefont {Goga},
  \citenamefont {Rzepiela}, \citenamefont {De~Vries}, \citenamefont {Marrink},\
  and\ \citenamefont {Berendsen}}]{goga2012efficient}%
  \BibitemOpen
  \bibfield  {author} {\bibinfo {author} {\bibfnamefont {N.}~\bibnamefont
  {Goga}}, \bibinfo {author} {\bibfnamefont {A.}~\bibnamefont {Rzepiela}},
  \bibinfo {author} {\bibfnamefont {A.}~\bibnamefont {De~Vries}}, \bibinfo
  {author} {\bibfnamefont {S.}~\bibnamefont {Marrink}}, \ and\ \bibinfo
  {author} {\bibfnamefont {H.}~\bibnamefont {Berendsen}},\ }\bibfield  {title}
  {\enquote {\bibinfo {title} {Efficient algorithms for {Langevin} and
  {D}{P}{D} dynamics},}\ }\href@noop {} {\bibfield  {journal} {\bibinfo
  {journal} {Journal of Chemical Theory and Computation}\ }\textbf {\bibinfo
  {volume} {8}},\ \bibinfo {pages} {3637--3649} (\bibinfo {year}
  {2012})}\BibitemShut {NoStop}%
\bibitem [{\citenamefont {Abraham}\ \emph {et~al.}(2015)\citenamefont
  {Abraham}, \citenamefont {Murtola}, \citenamefont {Schulz}, \citenamefont
  {P{\'{a}}ll}, \citenamefont {Smith}, \citenamefont {Hess},\ and\
  \citenamefont {Lindah}}]{Abraham2015}%
  \BibitemOpen
  \bibfield  {author} {\bibinfo {author} {\bibfnamefont {M.~J.}\ \bibnamefont
  {Abraham}}, \bibinfo {author} {\bibfnamefont {T.}~\bibnamefont {Murtola}},
  \bibinfo {author} {\bibfnamefont {R.}~\bibnamefont {Schulz}}, \bibinfo
  {author} {\bibfnamefont {S.}~\bibnamefont {P{\'{a}}ll}}, \bibinfo {author}
  {\bibfnamefont {J.~C.}\ \bibnamefont {Smith}}, \bibinfo {author}
  {\bibfnamefont {B.}~\bibnamefont {Hess}}, \ and\ \bibinfo {author}
  {\bibfnamefont {E.}~\bibnamefont {Lindah}},\ }\bibfield  {title} {\enquote
  {\bibinfo {title} {{GROMACS}: High performance molecular simulations through
  multi-level parallelism from laptops to supercomputers},}\ }\href {\doibase
  10.1016/j.softx.2015.06.001} {\bibfield  {journal} {\bibinfo  {journal}
  {SoftwareX}\ }\textbf {\bibinfo {volume} {1-2}},\ \bibinfo {pages} {19--25}
  (\bibinfo {year} {2015})}\BibitemShut {NoStop}%
\bibitem [{\citenamefont {Bonomi}\ \emph {et~al.}(2009)\citenamefont {Bonomi},
  \citenamefont {Branduardi}, \citenamefont {Bussi}, \citenamefont {Camilloni},
  \citenamefont {Provasi}, \citenamefont {Raiteri}, \citenamefont {Donadio},
  \citenamefont {Marinelli}, \citenamefont {Pietrucci}, \citenamefont
  {Broglia},\ and\ \citenamefont {Parrinello}}]{Bonomi2009}%
  \BibitemOpen
  \bibfield  {author} {\bibinfo {author} {\bibfnamefont {M.}~\bibnamefont
  {Bonomi}}, \bibinfo {author} {\bibfnamefont {D.}~\bibnamefont {Branduardi}},
  \bibinfo {author} {\bibfnamefont {G.}~\bibnamefont {Bussi}}, \bibinfo
  {author} {\bibfnamefont {C.}~\bibnamefont {Camilloni}}, \bibinfo {author}
  {\bibfnamefont {D.}~\bibnamefont {Provasi}}, \bibinfo {author} {\bibfnamefont
  {P.}~\bibnamefont {Raiteri}}, \bibinfo {author} {\bibfnamefont
  {D.}~\bibnamefont {Donadio}}, \bibinfo {author} {\bibfnamefont
  {F.}~\bibnamefont {Marinelli}}, \bibinfo {author} {\bibfnamefont
  {F.}~\bibnamefont {Pietrucci}}, \bibinfo {author} {\bibfnamefont {R.~A.}\
  \bibnamefont {Broglia}}, \ and\ \bibinfo {author} {\bibfnamefont
  {M.}~\bibnamefont {Parrinello}},\ }\bibfield  {title} {\enquote {\bibinfo
  {title} {{PLUMED: A} portable plugin for free-energy calculations with
  molecular dynamics},}\ }\href {\doibase 10.1016/j.cpc.2009.05.011} {\bibfield
   {journal} {\bibinfo  {journal} {Computer Physics Communications}\ }\textbf
  {\bibinfo {volume} {180}},\ \bibinfo {pages} {1961--1972} (\bibinfo {year}
  {2009})}\BibitemShut {NoStop}%
\bibitem [{\citenamefont {Tribello}\ \emph {et~al.}(2014)\citenamefont
  {Tribello}, \citenamefont {Bonomi}, \citenamefont {Branduardi}, \citenamefont
  {Camilloni},\ and\ \citenamefont {Bussi}}]{Tribello2014}%
  \BibitemOpen
  \bibfield  {author} {\bibinfo {author} {\bibfnamefont {G.~A.}\ \bibnamefont
  {Tribello}}, \bibinfo {author} {\bibfnamefont {M.}~\bibnamefont {Bonomi}},
  \bibinfo {author} {\bibfnamefont {D.}~\bibnamefont {Branduardi}}, \bibinfo
  {author} {\bibfnamefont {C.}~\bibnamefont {Camilloni}}, \ and\ \bibinfo
  {author} {\bibfnamefont {G.}~\bibnamefont {Bussi}},\ }\bibfield  {title}
  {\enquote {\bibinfo {title} {{PLUMED 2}: {New} feathers for an old bird},}\
  }\href {\doibase 10.1016/j.cpc.2013.09.018} {\bibfield  {journal} {\bibinfo
  {journal} {Computer Physics Communications}\ }\textbf {\bibinfo {volume}
  {185}},\ \bibinfo {pages} {604--613} (\bibinfo {year} {2014})}\BibitemShut
  {NoStop}%
\bibitem [{\citenamefont {Bonomi}\ \emph {et~al.}(2019)\citenamefont {Bonomi},
  \citenamefont {Bussi}, \citenamefont {Camilloni}, \citenamefont {Tribello},
  \citenamefont {Banáš}, \citenamefont {Barducci}, \citenamefont {Bernetti},
  \citenamefont {Bolhuis}, \citenamefont {Bottaro}, \citenamefont {Branduardi},
  \citenamefont {Capelli},\ and\ \citenamefont {Carloni}}]{PLUMED2019}%
  \BibitemOpen
  \bibfield  {author} {\bibinfo {author} {\bibfnamefont {M.}~\bibnamefont
  {Bonomi}}, \bibinfo {author} {\bibfnamefont {G.}~\bibnamefont {Bussi}},
  \bibinfo {author} {\bibfnamefont {C.}~\bibnamefont {Camilloni}}, \bibinfo
  {author} {\bibfnamefont {G.~A.}\ \bibnamefont {Tribello}}, \bibinfo {author}
  {\bibfnamefont {P.}~\bibnamefont {Banáš}}, \bibinfo {author} {\bibfnamefont
  {A.}~\bibnamefont {Barducci}}, \bibinfo {author} {\bibfnamefont
  {M.}~\bibnamefont {Bernetti}}, \bibinfo {author} {\bibfnamefont {P.~G.}\
  \bibnamefont {Bolhuis}}, \bibinfo {author} {\bibfnamefont {S.}~\bibnamefont
  {Bottaro}}, \bibinfo {author} {\bibfnamefont {D.}~\bibnamefont {Branduardi}},
  \bibinfo {author} {\bibfnamefont {R.}~\bibnamefont {Capelli}}, \ and\
  \bibinfo {author} {\bibfnamefont {P.}~\bibnamefont {Carloni}},\ }\bibfield
  {title} {\enquote {\bibinfo {title} {Promoting transparency and
  reproducibility in enhanced molecular simulations},}\ }\href {\doibase
  10.1038/s41592-019-0506-8} {\bibfield  {journal} {\bibinfo  {journal} {Nature
  Methods}\ }\textbf {\bibinfo {volume} {16}},\ \bibinfo {pages} {670--673}
  (\bibinfo {year} {2019})}\BibitemShut {NoStop}%
\bibitem [{\citenamefont {Bolhuis}, \citenamefont {Dellago},\ and\
  \citenamefont {Chandler}(2000)}]{bolhuis2000reaction}%
  \BibitemOpen
  \bibfield  {author} {\bibinfo {author} {\bibfnamefont {P.~G.}\ \bibnamefont
  {Bolhuis}}, \bibinfo {author} {\bibfnamefont {C.}~\bibnamefont {Dellago}}, \
  and\ \bibinfo {author} {\bibfnamefont {D.}~\bibnamefont {Chandler}},\
  }\bibfield  {title} {\enquote {\bibinfo {title} {Reaction coordinates of
  biomolecular isomerization},}\ }\href@noop {} {\bibfield  {journal} {\bibinfo
   {journal} {Proceedings of the National Academy of Sciences}\ }\textbf
  {\bibinfo {volume} {97}},\ \bibinfo {pages} {5877--5882} (\bibinfo {year}
  {2000})}\BibitemShut {NoStop}%
\bibitem [{\citenamefont {Dickson}\ \emph {et~al.}(2011)\citenamefont
  {Dickson}, \citenamefont {Maienschein-Cline}, \citenamefont {Tovo-Dwyer},
  \citenamefont {Hammond},\ and\ \citenamefont {Dinner}}]{dickson2011flow}%
  \BibitemOpen
  \bibfield  {author} {\bibinfo {author} {\bibfnamefont {A.}~\bibnamefont
  {Dickson}}, \bibinfo {author} {\bibfnamefont {M.}~\bibnamefont
  {Maienschein-Cline}}, \bibinfo {author} {\bibfnamefont {A.}~\bibnamefont
  {Tovo-Dwyer}}, \bibinfo {author} {\bibfnamefont {J.~R.}\ \bibnamefont
  {Hammond}}, \ and\ \bibinfo {author} {\bibfnamefont {A.~R.}\ \bibnamefont
  {Dinner}},\ }\bibfield  {title} {\enquote {\bibinfo {title} {Flow-dependent
  unfolding and refolding of an {R}{N}{A} by nonequilibrium umbrella
  sampling},}\ }\href@noop {} {\bibfield  {journal} {\bibinfo  {journal}
  {Journal of Chemical Theory and Computation}\ }\textbf {\bibinfo {volume}
  {7}},\ \bibinfo {pages} {2710--2720} (\bibinfo {year} {2011})}\BibitemShut
  {NoStop}%
\end{thebibliography}


\begin{widetext}

\appendix
\section{Formulas for the reactive current}
\label{TPT}

In this section, we derive an expression for the reactive current of a diffusion and use it to motivate %\eqref{JABtheta}
(7).  To this end, we introduce the generator, which describes the evolution of the expectation of functions over an infinitesimal time:
\begin{equation}\label{genexpdef}
    Lf(x) = \lim_{\tau\to 0}\frac{{\bf E}[f(X(\tau))|X(0)=x]-f(x)}{\tau}.
\end{equation}
%Because we also need steady-state backward-in-time statistics, we define the $\pi$-weighted adjoint $L_\pi^\dagger$ of $L$:
%\begin{equation}
   % \langle f,Lg\rangle = \langle L_\pi^\dagger f, g\rangle
%\end{equation}
%with inner product 
%\begin{equation}
%     \langle f,g\rangle= \int f(x)g(x)\pi(dx).
%\end{equation}
Backward-in-time statistics such as $q_-(x)$ can be accessed through the adjoint of $L$ in a  $\pi$-weighted inner product (see \cite{strahan2021long} for further discussion), and in this sense we can write
\begin{equation}\label{bwdgenexpdef}
   L_\pi^\dagger f(x)=\lim_{\tau\to 0}\frac{{\bf E}[f(X(-\tau))|X(0)=x]-f(x)}{\tau}.
\end{equation}
For the diffusive dynamics 
\begin{equation}
   \dot{x}(t) = b(x(t))+\sqrt{2}\sigma(x(t))\eta(t)
\end{equation}
with $a=\sigma\sigma^\top$ and $\eta$ is a white noise (i.e., the $\pi$-weighted inner product is $\langle\eta_i(t)\eta_j(s)\rangle = \delta_{ij}\delta(t-s)$),
these operators take the forms
\begin{align}\label{gendef}
{L}f(x) &= \sum_j b_j(x) \frac{\partial f}{\partial x_j}(x)
+ \sum_{ij} a_{ij}(x) \frac{\partial^2 f}{\partial x_i \partial x_j}(x)\\
L^{\dagger}_{\pi}f(x)&=-\sum_ib_i(x)\frac{\partial f(x)}{\partial x_i}+\frac{2}{\pi(x)}\sum_j\frac{\partial}{\partial x_i}(a_{ij}(x)\pi(x))\frac{\partial f(x)}{\partial x_i}+\sum_{ij}a_{ij}(x)\frac{\partial^2f(x)}{\partial x_i \partial x_j}.\label{bwdgendef}
\end{align}

\subsection{Reactive current for diffusive dynamics}

Here, we use the forms in \eqref{gendef} and \eqref{bwdgendef} to show 
\begin{equation}\label{JABinL}
\edits{I}_{AB}\cdot \nabla \theta(x)=
\frac{\pi(x)}{2}\left(q_-(x)\mathcal{L}[q_+\theta](x)-q_+(x)\mathcal{L}_{\pi}^{\dagger}[q_-\theta](x)\right)
\end{equation}
for $x\in (A\cup B)^c$.  For clarity, we suppress the $x$ arguments of functions in the rest of this section.  

We start from the $i$-th component of the reactive current \cite{Vanden-Eijnden2006}:
\begin{equation}\label{Jabdef}
\edits{I}_{AB,i} = q_+q_-\edits{I}_i+\pi\bigg(q_-\sum_ja_{ij}\frac{\partial q_+}{\partial x_j}-q_+\sum_ja_{ij}\frac{\partial q_-}{\partial x_j}\bigg)
\end{equation}
where $\edits{I}_i$ is the $i$-th component of the equilibrium probability current: 
\begin{equation} \label{Jdef}
\edits{I}_i = \pi b_i-\sum_j\frac{\partial}{\partial x_j}\big(\pi a_{ij}\big).
\end{equation}
Substituting \eqref{Jdef} into \eqref{Jabdef}, summing over components, rearranging, and dotting with $\nabla\theta$,
\begin{align}
\edits{I}_{AB}\cdot\nabla \theta =\frac{\pi}{2}\sum_i \Bigg[ q_-  \bigg( &q_+b_i-\frac{1}{\pi}\sum_j q_+\frac{\partial}{\partial x_j}\big(\pi a_{ij}\big)+2\sum_ja_{ij}\frac{\partial q_+}{\partial x_j}\bigg) \nonumber \\
+q_+  \bigg(  &q_-b_i-\frac{1}{\pi}\sum_j q_-\frac{\partial}{\partial x_j}\big(\pi a_{ij}\big)- 2\sum_ja_{ij}\frac{\partial q_-}{\partial x_j}\bigg)\Bigg]\frac{\partial \theta}{\partial x_i}.\label{eq:JABdeltheta}
\end{align}
Adding the identity  $\theta L q^+=0$ to the first set of parentheses and rearranging, 
\begin{multline}
%\sum_i\bigg(q_+b_i-\frac{1}{\pi} q_+ \sum_j \frac{\partial}{\partial x_j}\big(\pi a_{ij}\big)+2\sum_ja_{ij}\frac{\partial q_+}{\partial x_j}\bigg)\frac{\partial \theta}{\partial x_i}
%+ \sum_i\theta\bigg(b_i \frac{\partial q_+}{\partial x_j}
%+ \sum_{j} a_{ij} \frac{\partial^2 q_+}{\partial x_i \partial x_j}\bigg)\\
%= 
\sum_i b_i\frac{\partial q_+ \theta}{d x_i}+\sum_{i,j} a_{ij} \frac{\partial^2 q_+ \theta}{\partial x_i \partial x_j}-\frac{1}{\pi} q_+ \sum_{i,j} \frac{\partial}{\partial x_j}\big(\pi a_{ij}\big)\frac{\partial \theta}{\partial x_i}+q_+\sum_{i,j}a_{ij}\frac{\partial^2 \theta}{\partial x_i \partial x_j}\\
=L[q_+ \theta]-\frac{1}{\pi} q_+ \sum_j \frac{\partial}{\partial x_j}\big(\pi a_{ij}\big)\frac{\partial \theta}{\partial x_i}+q_+\sum_{i,j}a_{ij}\frac{\partial^2 \theta}{\partial x_i \partial x_j}.
\end{multline}
Similarly, adding the identity $\theta L^\dagger_{\pi}q_-=0$  to the second set of parentheses in \eqref{eq:JABdeltheta}
and rearranging, 
\begin{multline}
\sum_i\bigg(q_-b_i-\frac{1}{\pi}\sum_j q_-\frac{\partial}{\partial x_j}\big(\pi a_{ij}\big)- 2\sum_ja_{ij}\frac{\partial q_-}{\partial x_j}\bigg)\frac{\partial \theta}{\partial x_i}\\
+\sum_i\theta\bigg( b_i\frac{\partial q_-}{\partial x_i}-\frac{2}{\pi}\frac{\partial}{\partial x_i}(a_{ij}\pi)\frac{\partial q_-}{\partial x_i}-\sum_{j}a_{ij}\frac{\partial^2q_-}{\partial x_i \partial x_j}\bigg)\\
=-L_{\pi, i}^\dagger[q_-\theta]+\frac{1}{\pi} q_- \sum_j \frac{\partial}{\partial x_j}\big(\pi a_{ij}\big)\frac{\partial \theta}{\partial x_i}-q_-\sum_{i,j}a_{ij}\frac{\partial^2 \theta}{\partial x_i \partial x_j}.
\end{multline}
Substituting these expressions into \eqref{eq:JABdeltheta} gives \eqref{JABinL}.

\subsection{General form for the reactive current}

Given \eqref{JABinL}, we can use \eqref{genexpdef} and \eqref{bwdgenexpdef} to write
\begin{align}
\edits{I}_{AB}\cdot \nabla \theta(x) = \lim_{\tau\to 0}\frac{\pi(x)}{2\tau}
%\frac{\pi(x)}{2}\bigg(q_-(x)\lim_{\tau\to 0}\frac{{\bf E}[\theta(X(\tau))q_+(X(\tau))|X(0)=x]-\theta(x)q_+(x)}{\tau} \\
%-&q_+(x)\lim_{\tau\to 0}\frac{{\bf E}[\theta(X(-\tau))q_-(X(-\tau))|X(0)=x]-\theta(x)q_-(x)}{\tau}\bigg)\\
\{&q_-(x) {\bf E}[\theta(X(\tau))q_+(X(\tau))|X(0)=x]\nonumber\\-&q_+(x){\bf E}[\theta(X(-\tau))q_-(X(-\tau))|X(0)=x]\}.\label{JABexplicit}
\end{align}
We can re-write the first term on the right hand side as
\begin{align}
{\bf E}[\theta(X(\tau))q_+(X(\tau))|X(0)=x]
&={\bf E}[{\bf E} [ \theta(X(\tau)) \mathbf{1}_B(X(t_+(0)))\, |\, X(\tau),\, X(0)]|X(0)=x]\\
&={\bf E}[ \theta(X(\tau)) \mathbf{1}_B(X(t_+(0)))\, |\, X(0)=x]\\
&=q_+(x){\bf E}[\theta(X(\tau))\, |\, X(t_+(0))\in B,\  X(0)=x].
\end{align}
The second equality follows from the tower property of conditional expectations.  The third equality follows from the definition of the conditional probability %as the ratio of the joint probability and the probability of the condition, %${\bf E}[\theta(X(\tau))|X(t_+(0))\in B,\  X(0)=x]$
and the definition of the committor, $q_+(x)={\bf P}[X(t_+(0))\in B | X(0)=x ]$.
Applying the same logic to the backward-in-time process,
\begin{equation}
{\bf E}[\theta(X(-\tau))q_-(X(-\tau))|X(0)=x] 
%&= {\bf E}[\theta(X(-\tau)){\bf P}[X(t_-(0))\in A | X(0)=X(-\tau) ]| X(0)=x]\\
%&={\bf P}[X(t_-(0))\in A | X(0)=x ]{\bf E}[\theta(X(-\tau))|X(t_-(0))\in A , X(0)=x]\\
=q_-(x){\bf E}[\theta(X(-\tau))|X(t_-(0))\in A, X(0)=x].
\end{equation}
Substituting into \eqref{JABexplicit} and combining terms, 
%
%\begin{align}
%\edits{I}_{AB}\cdot \nabla \theta(x)= \pi(x) q_-(x)q_+(x)\lim_{\tau\to 0} \frac{1}{2\tau}\{ &{\bf E}[\theta(X(\tau))|X(t_+(0))\in B, X(0)=x]\nonumber\\-&{\bf E}[\theta(X(-\tau))|X(t_-(0))\in A,  X(0)=x]\}
%\end{align}
\begin{multline}
\edits{I}_{AB}\cdot \nabla \theta(x)=\pi(x) q_-(x)q_+(x)\\ \times
\lim_{\tau\to 0}\frac{ {\bf E}[(\theta(X(\tau))-\theta(X(-\tau))\, |\, 
X(t_-(0))\in A,\, X(t_+(0))\in B,\, 
X(0)=x]}{2\tau}
\end{multline}
for $x\in (A\cup B)^c$.  In combining terms into a single expectation, we have exploited the fact that, for a fixed $X(0)$ of a Markov process, $X(-\tau)$ is independent of  $X(t_+(0))$, and $X(\tau)$ is independent of  $X(t_-(0))$.

%\begin{equation}
%\edits{I}_{AB}\cdot\nabla\theta(x)=\pi_{AB} \frac{ E_{AB}[(\theta(X(t+\tau))-\theta(X(t-\tau))|X(t)=x]}{2\tau}
%\end{equation}
%
%where we use $E_{AB}$ to denote an average over the distribution of reactive trajectories. 

\section{Forward and backward reaction quantities}

To quantify the convergence of our results, we compute the backward committor and plot $q^+-q^--1$, which should be zero for a reversible system.  We show these data as a function of the number of completed crossings between the metastable states in Figs.\ \ref{fig:commitdiffs1} and \ref{fig:commitdiffs2}.  We also plot the sum of the reactive currents from A to B and from B to A. Because the dynamics are underdamped, this quantity should not be exactly zero, though we would expect it to be close in general.  We see that this is the case in Fig.\ \ref{fig:currentdiff}, with the main deviations close to the boundaries of the metastable states.

\begin{figure}
\begin{tabular}{cc}
 \includegraphics[width=65mm]{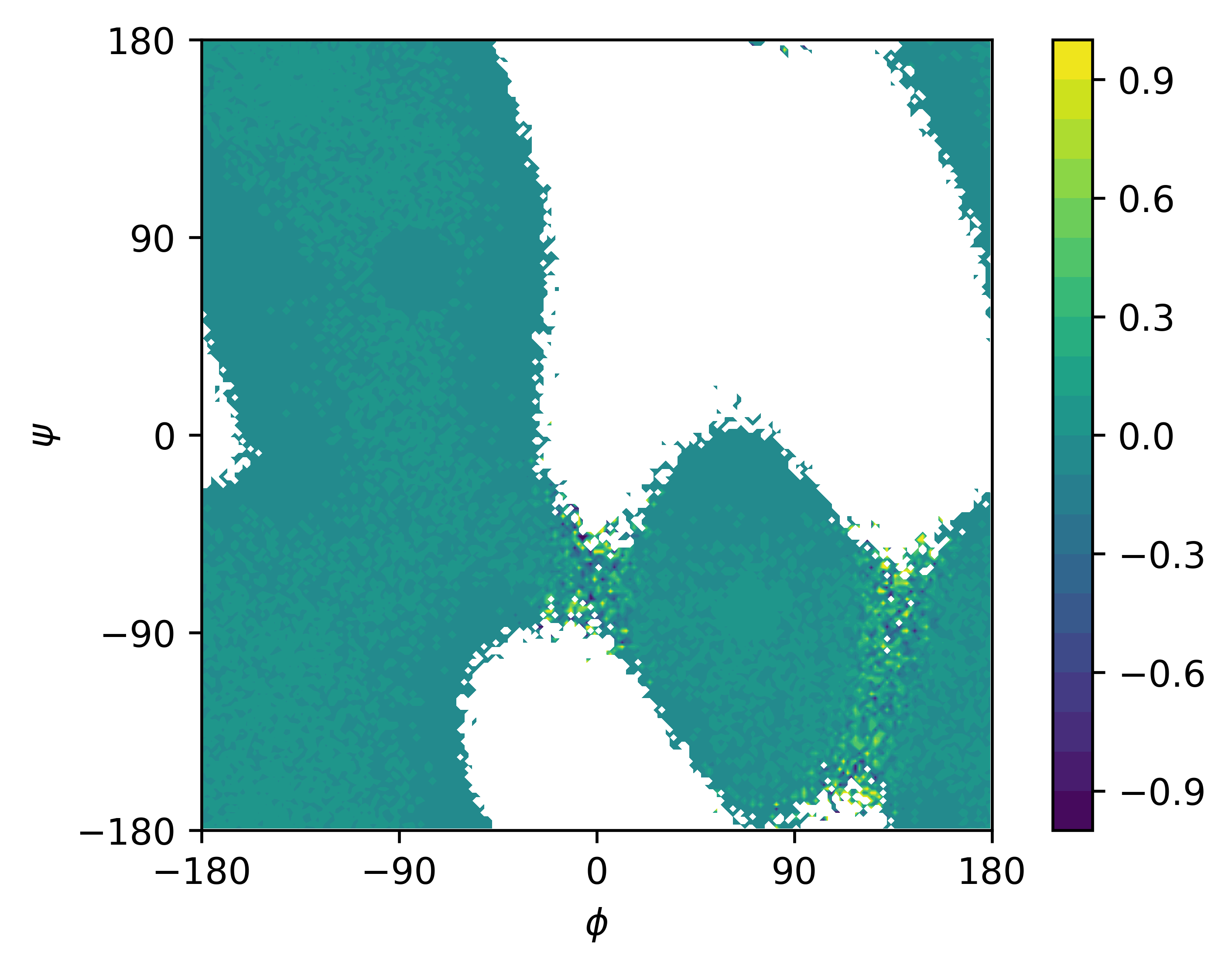} &   \includegraphics[width=65mm]{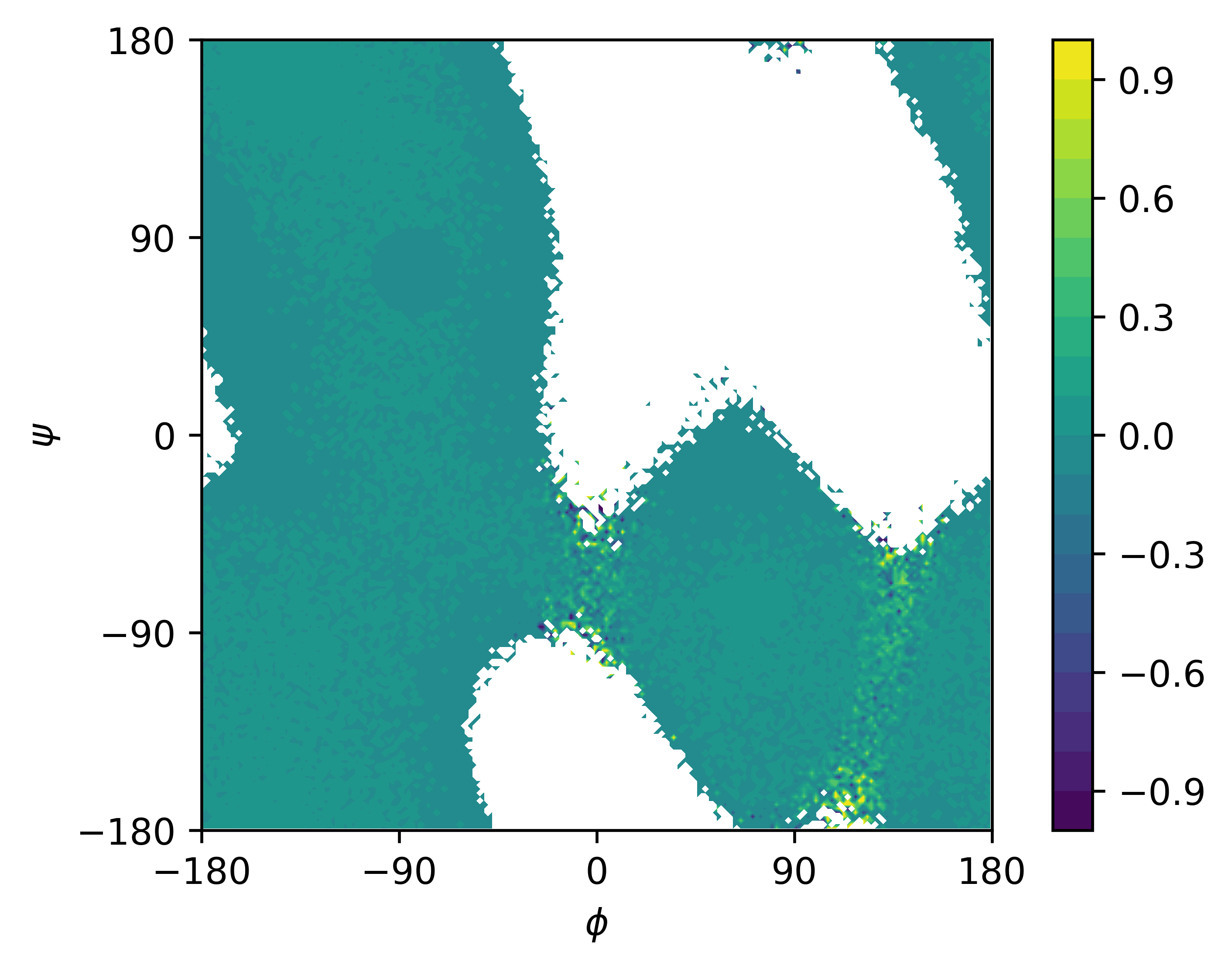} \\
(a)  & (b)  \\[6pt]
 \includegraphics[width=65mm]{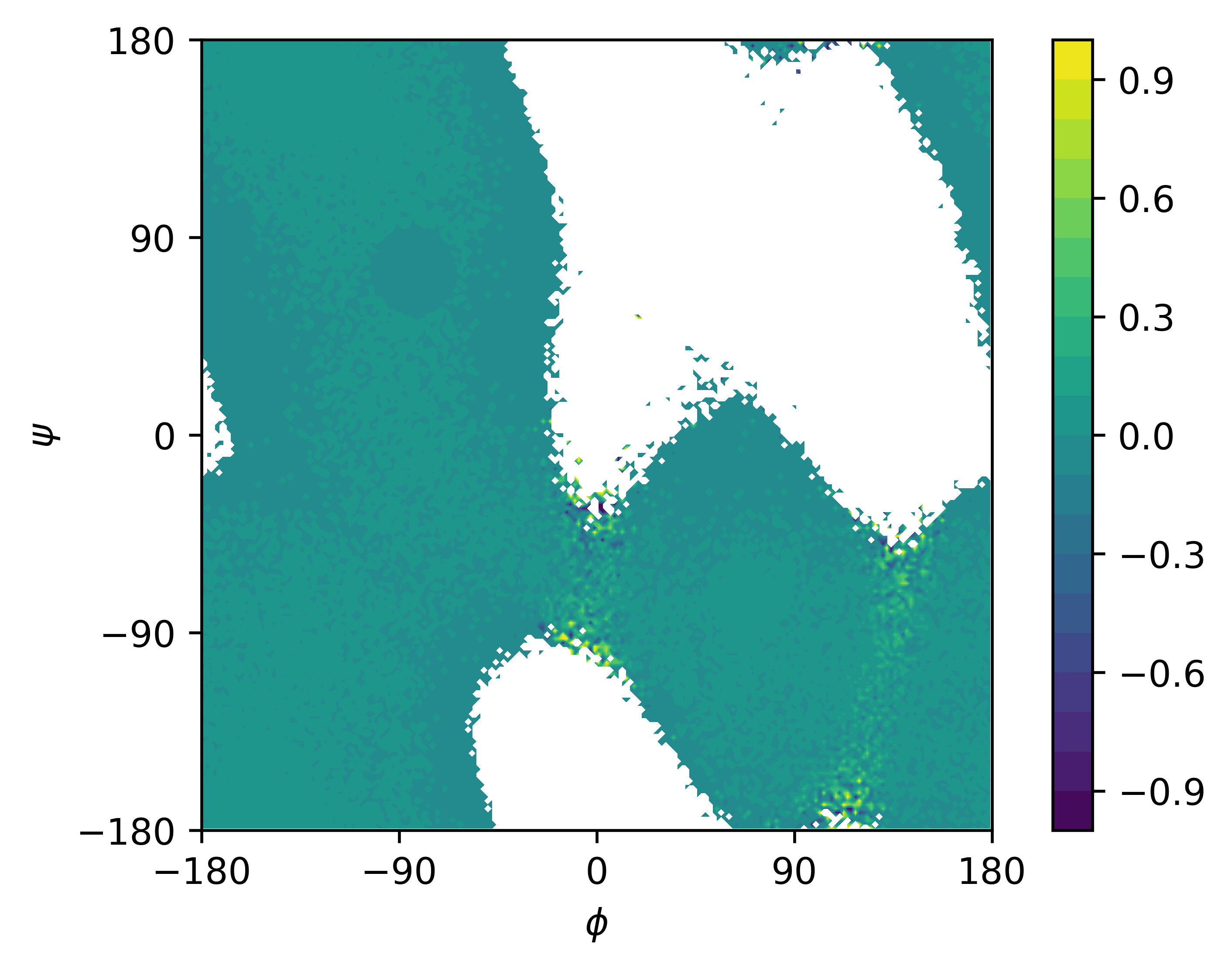} &   \includegraphics[width=65mm]{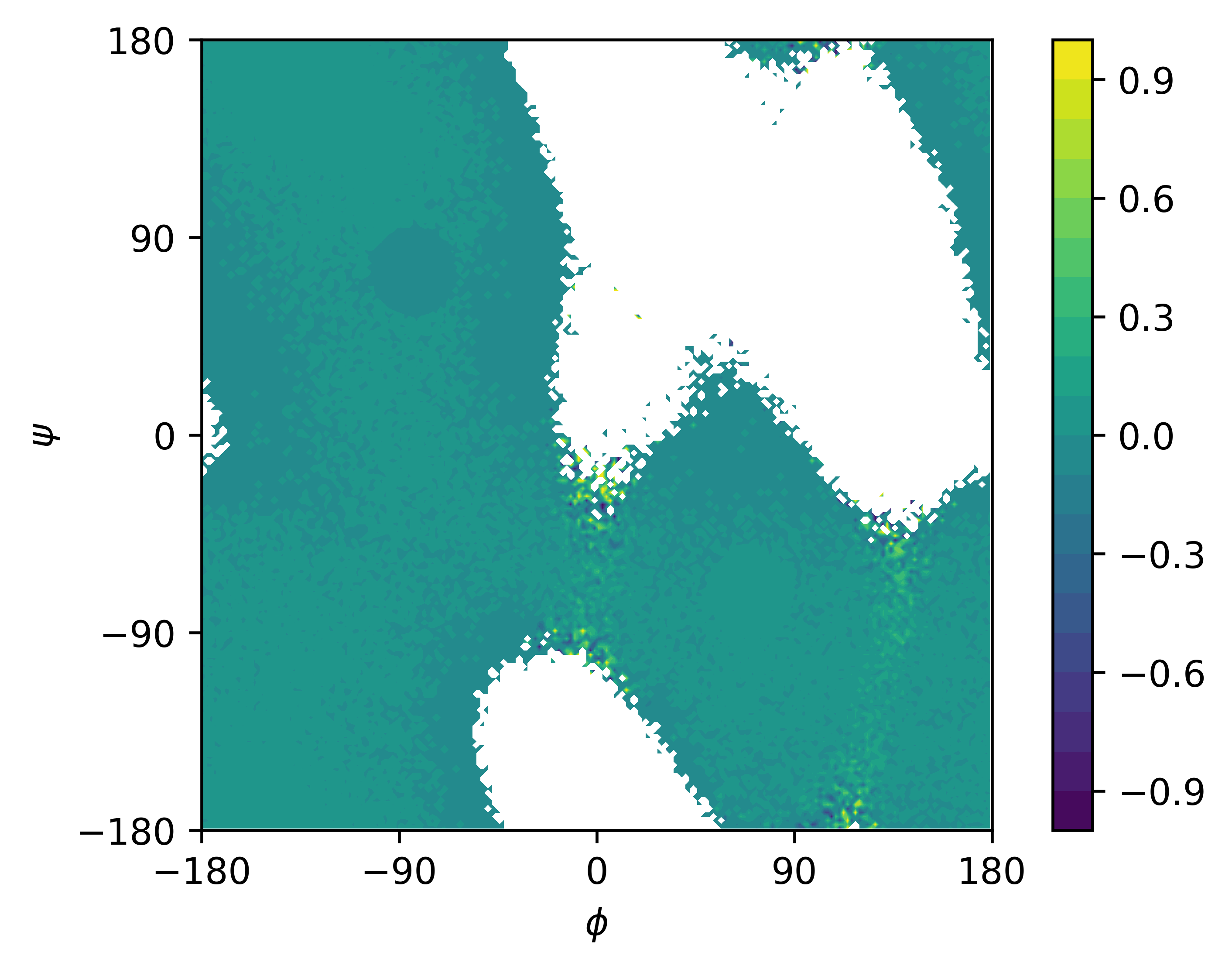} \\
(c)  & (d)  \\[6pt]
\end{tabular}
\caption{Sum of forward committor and backward committors offset by 1 (i.e., $q^++q^--1$) projected onto $\phi$ and $\psi$ after harvesting (a) 100, (b) 250, (c) 500, (d) 1000 metastable state crossings.\label{fig:commitdiffs1}}
\end{figure}

\begin{figure}
\begin{tabular}{cc}
  \includegraphics[width=65mm]{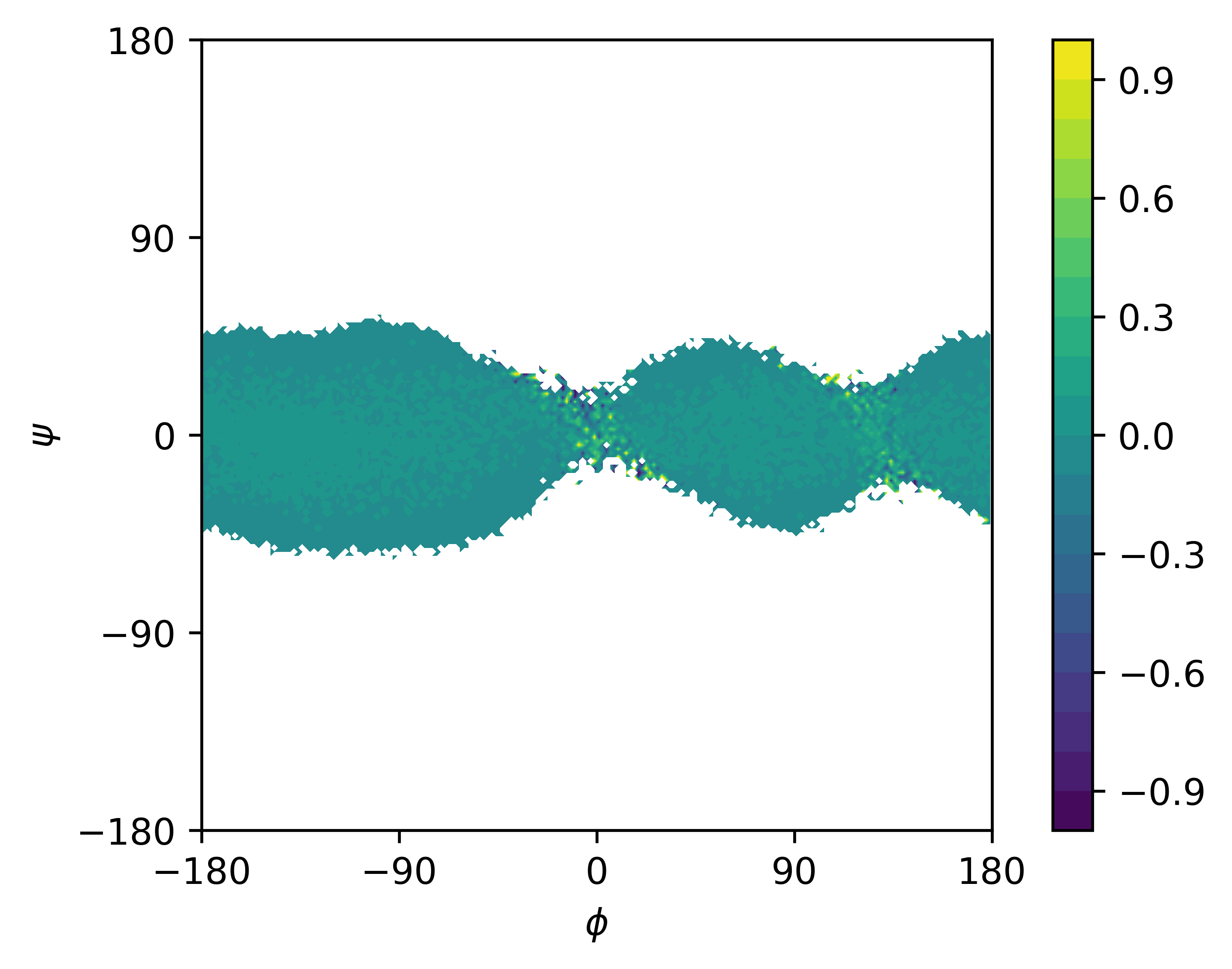} &   \includegraphics[width=65mm]{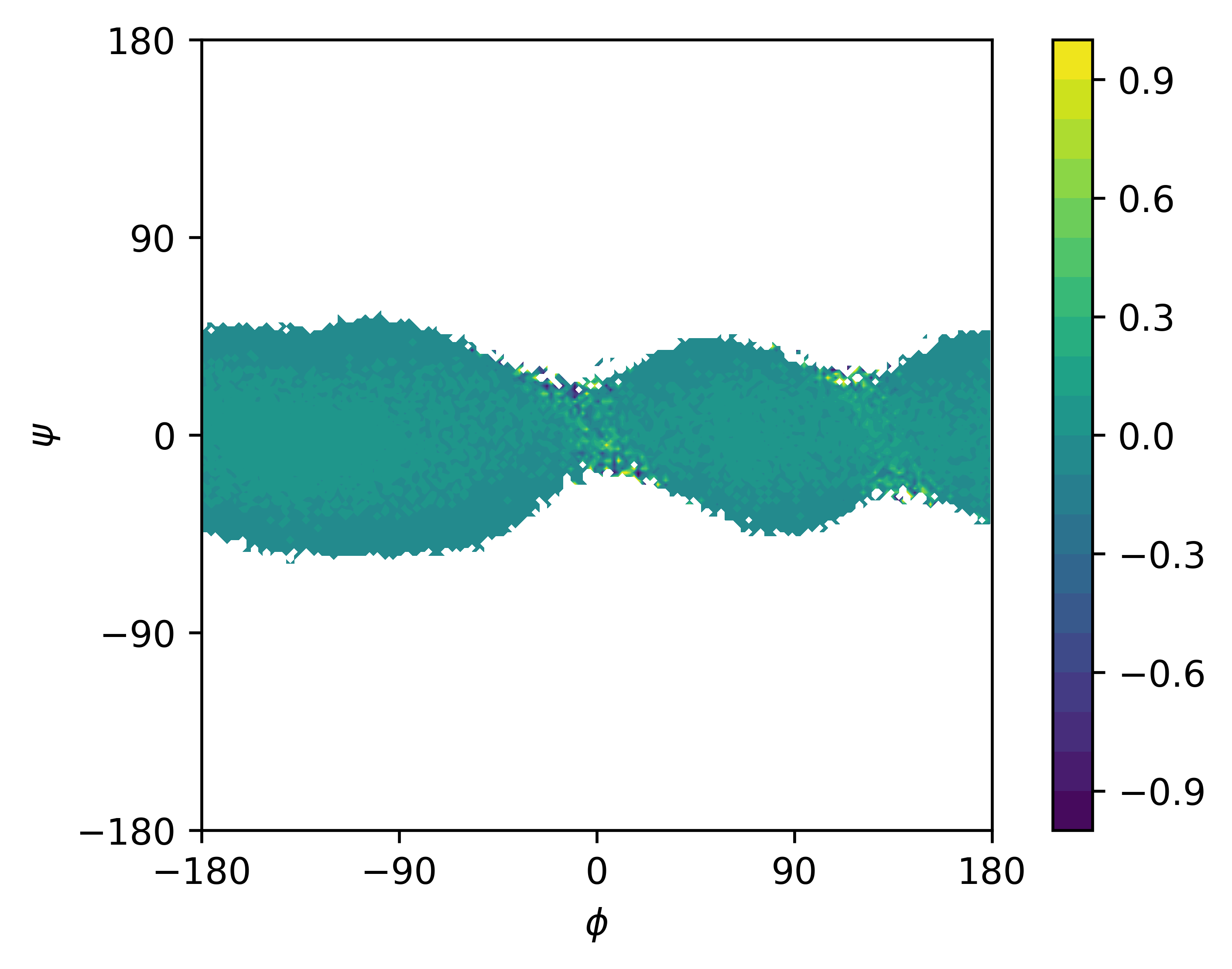} \\
(a)  & (b)  \\[6pt]
 \includegraphics[width=65mm]{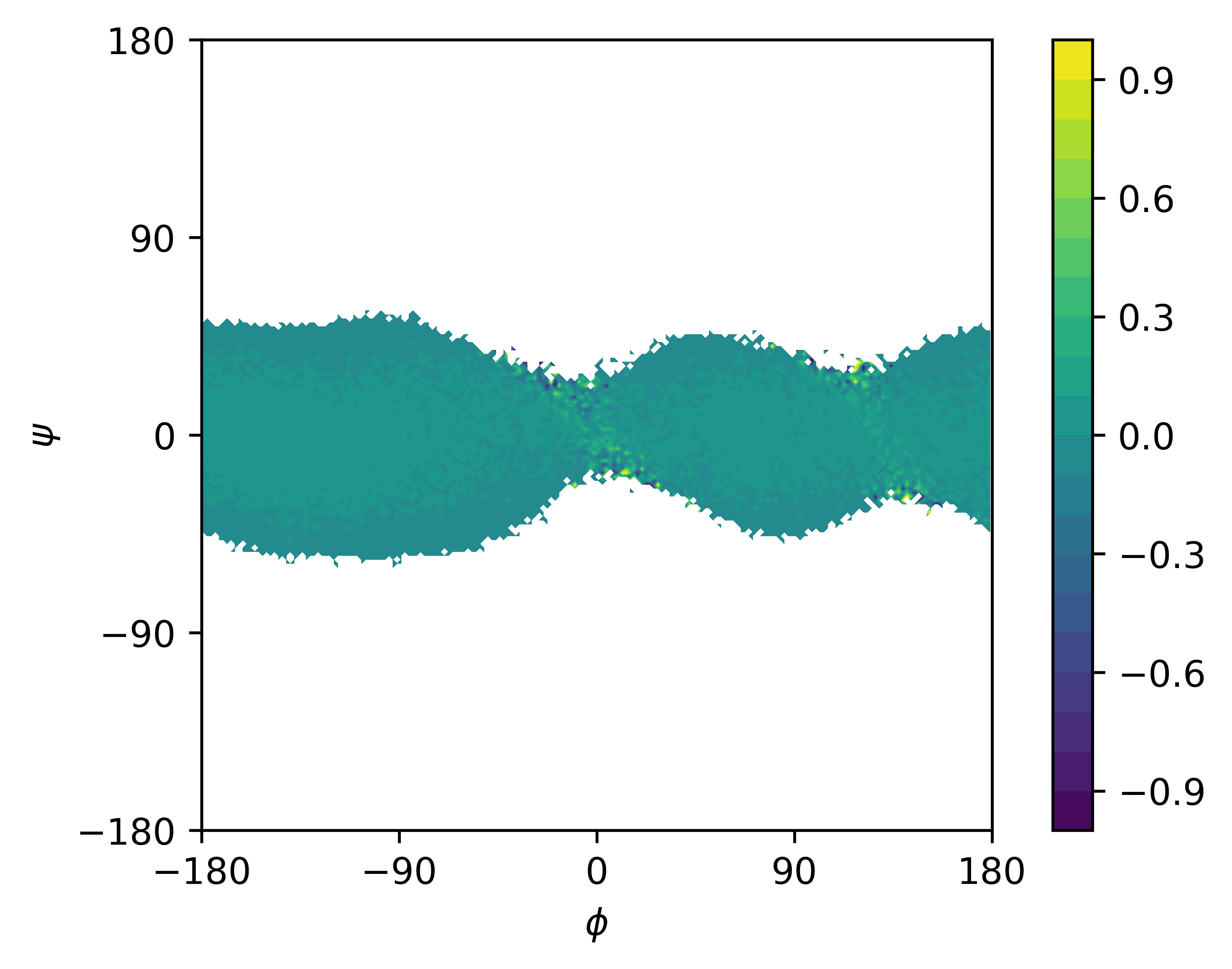} &   \includegraphics[width=65mm]{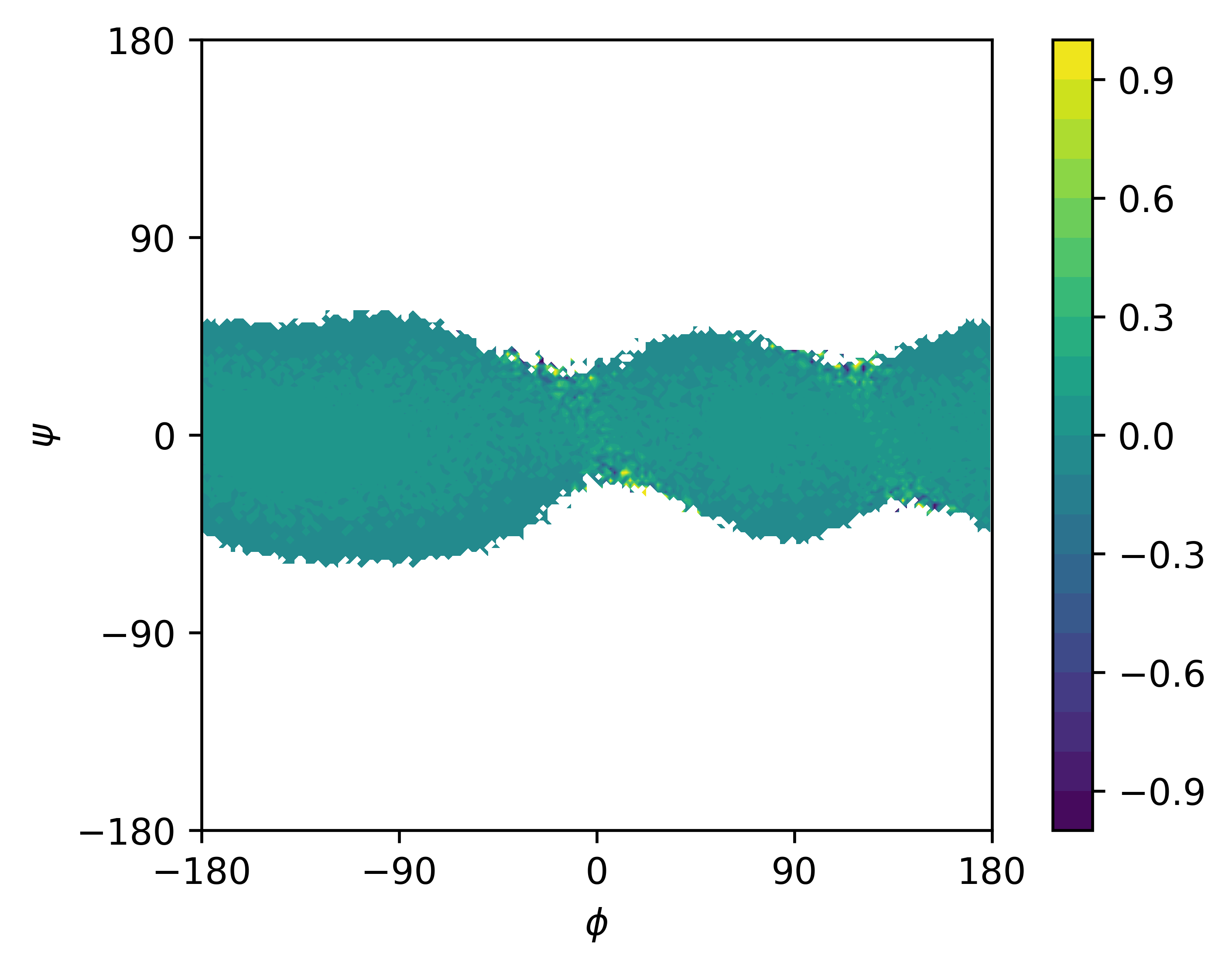} \\
(c)  & (d)  \\[6pt]
\end{tabular}
\caption{Sum of forward committor and backward committors offset by 1 (i.e., $q^++q^--1$) projected onto $\phi$ and $\omega$ after harvesting (a) 100, (b) 250, (c) 500, (d) 1000 metastable state crossings. \label{fig:commitdiffs2}}
\end{figure}

\begin{figure}[h!]
\centering
\includegraphics[scale=0.7]{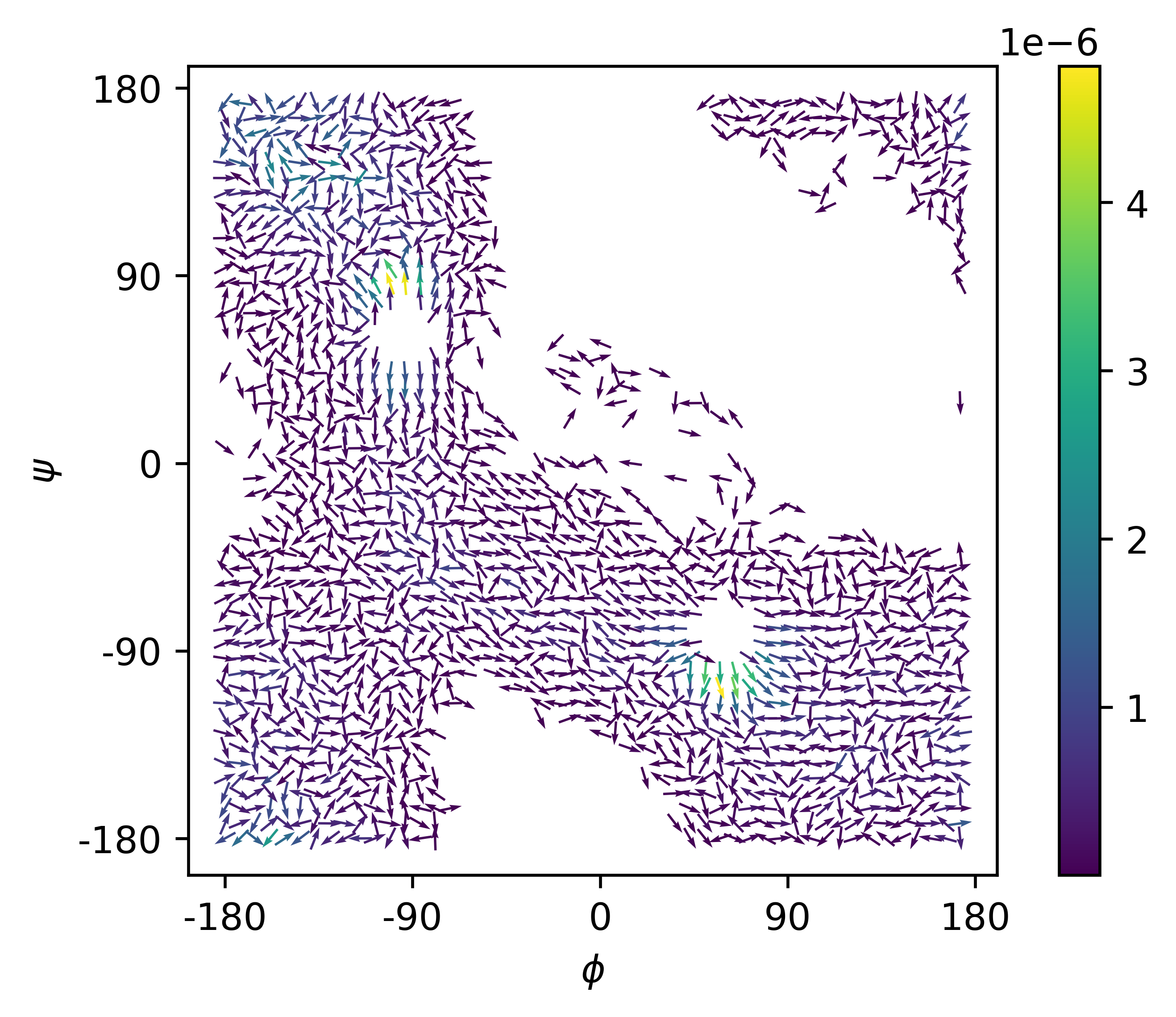}\\
\includegraphics[scale=0.7]{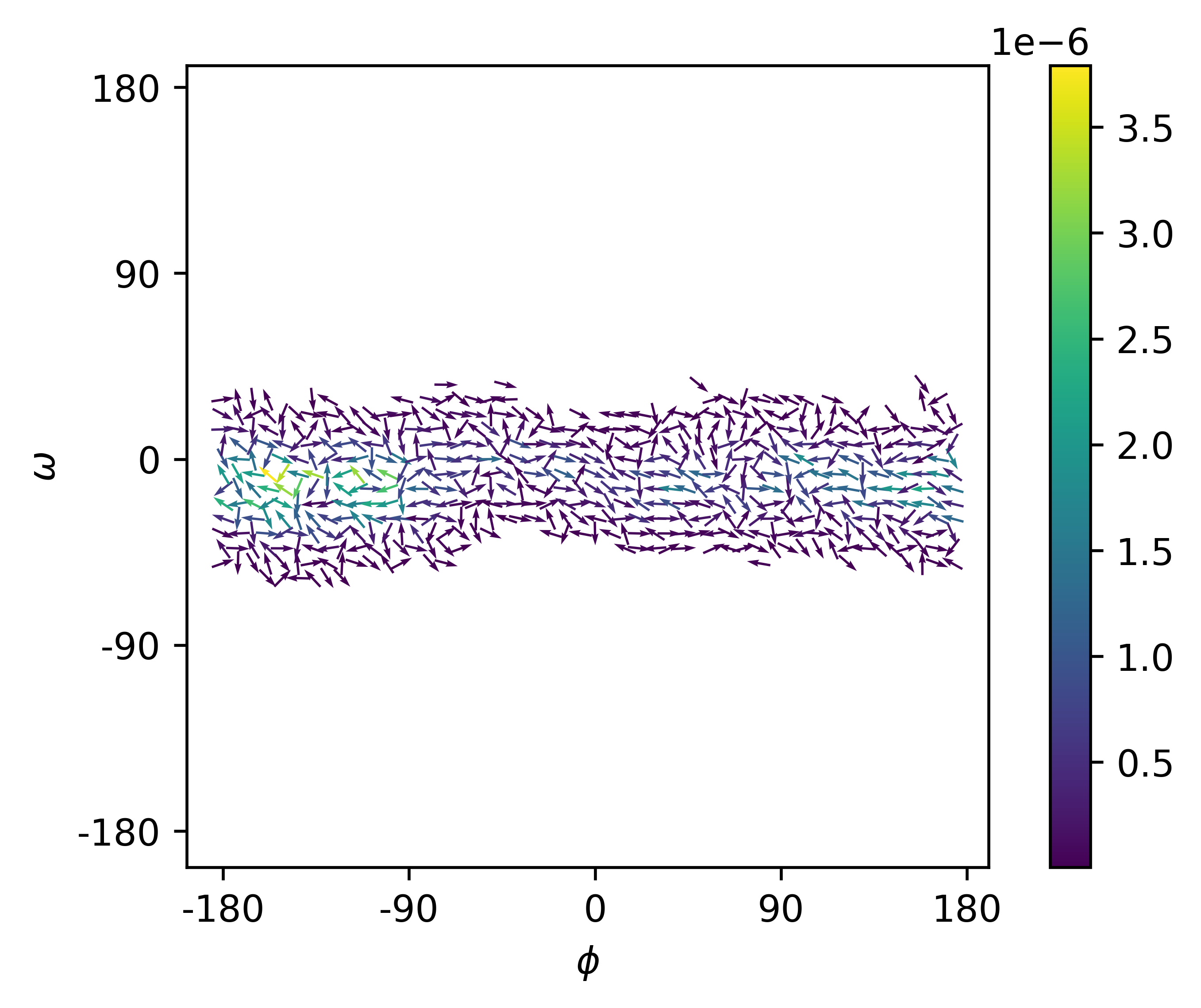}
\caption{ Sum of forward and backward reactive currents for indicated axes. \label{fig:currentdiff}
}
\end{figure}

\end{widetext}

\end{document}